\documentclass[preprint,10pt]{elsarticle}
\usepackage[top=2cm, bottom=2cm, left=2.5cm, right=3.5cm]{geometry}

\usepackage{amssymb}
\usepackage{amsmath}

\usepackage{lmodern}
\usepackage{ragged2e}
\usepackage{microtype}

\usepackage{algorithm}  
\usepackage{algorithmic}  

\usepackage{CJKutf8}



\makeatletter
\renewcommand{\maketag@@@}[1]{\hbox{\m@th\normalsize\normalfont#1}}%
\makeatother

\usepackage{changes}

\begin{document}
\begin{CJK}{UTF8}{gbsn}

\begin{frontmatter}



\title{A stochastic agent-based model for simulating tumor-immune dynamics and evaluating therapeutic strategies} %% Article title


\author[label1]{Yuhong Zhang(张玉红)\corref{coauthor}} %% Author name

\author[label2]{Chenghang Li(李城杭)\corref{coauthor}} %% Author name

\author[label2]{Boya Wang(王博雅)} %% Author name

\author[label2,label3]{Jinzhi Lei(雷锦志)\corref{cor1}} %% Author name
\cortext[coauthor]{These authors contributed equally to this work}
\cortext[cor1]{Corresponding author. \\ Email: jzlei@tiangong.edu.cn (J. Lei)}

\affiliation[label1]{organization={School of Software, Tiangong University},%Department and Organization 
            city={Tianjin},
            postcode={300387}, 
            country={China}}

\affiliation[label2]{organization={School of Mathematical Sciences, Tiangong University},%Department and Organization 
            city={Tianjin},
            postcode={300387}, 
            country={China}}

\affiliation[label3]{organization={Center for Applied Mathematics, Tiangong University},%Department and Organization 
            city={Tianjin},
            postcode={300387}, 
            country={China}}

%\textbf{Abstract:\ \ }
\begin{abstract}
Tumor-immune interactions are central to cancer progression and treatment outcomes. In this study, we present a stochastic agent-based model that integrates cellular heterogeneity, spatial cell-cell interactions, and drug resistance evolution to simulate tumor growth and immune response in a two-dimensional microenvironment. The model captures dynamic behaviors of four major cell types--tumor cells, cytotoxic T lymphocytes, helper T cells, and regulatory T cells--and incorporates key biological processes such as proliferation, apoptosis, migration, and immune regulation. Using this framework, we simulate tumor progression under different therapeutic interventions, including radiotherapy, targeted therapy, and immune checkpoint blockade. Our simulations reproduce emergent phenomena such as immune privilege and spatial immune exclusion. Quantitative analyses show that all therapies suppress tumor growth to varying degrees and reshape the tumor microenvironment. Notably, combination therapies--especially targeted therapy with immunotherapy--achieve the most effective tumor control and delay the emergence of resistance. Additionally, sensitivity analyses reveal a nonlinear relationship between treatment intensity and therapeutic efficacy, highlighting the existence of optimal dosing thresholds. This work demonstrates the utility of agent-based modeling in capturing complex tumor-immune dynamics and provides a computational platform for optimizing cancer treatment strategies. The model is extensible, biologically interpretable, and well-suited for future integration with experimental or clinical data.
\end{abstract}

%\textbf{Keyword:\ \ }
\end{frontmatter}

\section{Introduction}
\label{sec1}
Tumor-immune interactions are central to cancer progression, involving a complex dynamic equilibrium between tumor cells and the immune system \cite{Chen.Immunity.2013,deVisser.CancerCell.2023}. The adaptive immune system exerts a precise killing effect in antitumor immunity through antigen-specific recognition mechanisms \cite{Gajewski.NatImmunol.2013,Vesely.AnnuRevImmunol.2011}. Consequently, as the core force of immune responses, T cell-mediated cellular immunity enables tumor-specific clearance \cite{St-Paul.TrendsCellBiol.2020}. Conversely, tumor cells employ multidimensional immune evasion strategies to subvert antitumor immunity, primarily involving antigen presentation defects, immunosuppressive microenvironment formation, immunosuppressive cell recruitment, and immune checkpoint activation. These mechanisms collectively and systematically impair T cell-mediated specific immune clearance and promote tumor immune evasion \cite{Galassi.CancerCell.2024,Schreiber.Science.2011}. Quantitative study of tumor-immune interactions can help reveal tumor immune evasion mechanisms and support the development of more effective therapeutic strategies.

Agent-based modeling (ABM) is a stochastic computational modeling approach that decomposes systems into autonomous, heterogeneous agents, each interacting with the environment and other agents through predefined local rules and decision-making mechanisms. These microscopic interactions collectively give rise to macroscopic system features, enabling the quantitative study of the dynamic evolution of complex systems \cite{Albi.MathComputSimulat.2022,Metzcar.JCOClinCancerInform.2019,Deisboeck.AnnuRevBiomedEng.2011}. The ABM approach effectively captures system heterogeneity, stochasticity, and nonlinearity, making it particularly suitable for the study of multiscale complex systems and has been widely applied in computational and systems biology \cite{Merelli.BriefBioinform.2007}. Using ABM to model tumor-immune system interactions enables characterization of both single-cell heterogeneity and cell population-level behaviors and spatial features \cite{Albi.MathComputSimulat.2022,Anderson.Cell.2006}. It not only complements experimental studies effectively but also predicts novel experimental outcomes through stochastic simulations and reveals microscopic dynamics unobservable by conventional methods \cite{Anderson.NatRevCancer.2008,West.TRENDSCELLBIOL.2023}.

In recent years, significant advancements have been made in the application of ABM to tumor research, yielding successful studies on key processes including tumor evolution, immune interactions, and treatment responses \cite{Zhang.JMathBiol.2009}. Anderson et al. integrated partial differential equations with cellular automata to model the effects of extracellular matrix and oxygen concentration in the microenvironment on tumor evolutionary dynamics, revealing how microenvironmental selection pressures can drive phenotypic evolution and morphological changes \cite{Anderson.Cell.2006}. Zhang et al. employed ABM to characterize epidermal growth factor receptor-driven molecular networks, simulating glioma growth and invasion across multiple scales \cite{Zhang.MathComputSimul.2009}. Gong et al. developed a multiscale ABM incorporating heterogeneous immune checkpoint expression to simulate spatiotemporal tumor-immune interactions under immune checkpoint inhibitor therapy and predicted the impacts of mutation burden and antigen strength on treatment response \cite{Gong.JRSocInterface.2017}. Jalalimanesh et al. established a multiscale ABM to simulate vascularized tumor growth and radiotherapy response \cite{Jalalimanesh.MathComputSimul.2017}. In addition, Hickey et al. combined multi-source data with multiscale ABM to elucidate the critical role of tumor phenotypic switching in T cell therapy \cite{Hickey.CellSystems.2024}. While existing ABM studies have been applied to simulate complex tumor-immune interactions and treatment responses, they generally lack systematic integration of cellular heterogeneity and drug resistance mechanisms, which is the core focus of this study.

Tumor heterogeneity is a core feature of tumor tissues, manifested as diversity in genetic mutations, phenotypic characteristics, and functional behaviors \cite{Kashyap.TrendsBiotechnol.2022,Hausser.NatRevCancer.2020,Turajlic.NatRevGenet.2019}. This heterogeneity directly influences tumor proliferation, metastasis, and treatment response, especially by shaping the complex cellular interaction networks within the tumor microenvironment (TME), which significantly affect the efficacy of immunotherapy and the emergence of drug resistance. Tumor drug resistance remains one of the main challenges in cancer treatment, with complex and diverse underlying mechanisms, including drug target mutations, apoptosis inhibition, and therapy pressure-induced epigenetic alterations \cite{Marine.NatRevCancer.2020,Vasan.Nature.2019,Shi.SignalTransductTargetTher.2023,Zhang:2021kg}. Crucially, heterogeneity-driven clonal evolution promotes the selective expansion of drug-resistant subpopulations under therapeutic pressure, ultimately leading to treatment failure \cite{Gatenby.NatRevClinOncol.2020}. Therefore, deciphering cellular heterogeneity and drug resistance in tumor evolution is crucial for developing more effective therapeutic strategies.

In this study, we developed an ABM that incorporates cellular heterogeneity and tumor drug resistance to simulate the complex interactions among tumor cells and cytotoxic T cells (CTLs), helper T cells (Th cells), and regulatory T cells (Tregs) in the TME, and to characterize the spatiotemporal dynamic behavior of cells in two-dimensional space. For each cell type, we defined behavioral rules and mathematical formulations of cell state transition rates, including proliferation, death, and migration. The model generates high-resolution images of TME evolution during tumor growth and treatment. Within this framework, we simulated the spatiotemporal evolution and drug resistance dynamics of a heterogeneous tumor in the tumor-immune microenvironment under multiple treatment regimens and varying treatment intensities.

Simulation results revealed the complex dynamic equilibrium between tumor and immune cells under untreated conditions and identified the phenomenon of "immune privilege", in which tumors evade immune surveillance. We further contrasted different therapeutic strategies, demonstrating that radiotherapy suppressed tumor growth, reduced volume and cell density, attenuated Treg-mediated immunosuppression, enhanced CTL infiltration, and improved the TME. Targeted therapy effectively killed tumor cells and enhanced antitumor immunity by boosting T cell proliferation. Immunotherapy restricted tumor expansion, enhanced the immune response, and optimized the antitumor microenvironment by activating T cell function. Comparative analysis indicated that targeted therapy accelerated the selection of drug-resistant clones, whereas immunotherapy preserved some sensitive subpopulations, suggesting that combined strategies might delay the onset of drug resistance. Evaluations of different therapeutic strategies demonstrated that combination therapies yielded significantly superior efficacy compared to monotherapies, particularly radiotherapy combined with targeted therapy, which shortened treatment cycles and markedly improved therapeutic outcomes. Finally, quantitative analysis showed that increasing treatment intensity can improve antitumor efficacy, but an optimal threshold exists--suggesting that moderate intensification may enhance outcomes, while excessive treatment could disrupt immune microenvironment homeostasis.

The specific structure of this paper is as follows: Section \ref{Sec2} introduces the interaction mechanisms and state transition rates of the four cell types considered in the model, along with detailed descriptions of the design of the discrete stochastic simulator, computational workflow, and code architecture. Next, Section \ref{Sec3.1} depicts tumor growth under immune surveillance without treatment intervention. Section \ref{Sec3.2} investigates tumor response to therapy and tumor-immune system dynamics in the scenarios of radiotherapy, targeted therapy, and immunotherapy. Section \ref{Sec3.3} compares and analyzes the dynamic evolution of tumor cell drug resistance under targeted therapy and immunotherapy. Section \ref{Sec3.4} comprehensively evaluates the therapeutic efficacy of three monotherapies and combination therapies. Section \ref{Sec3.5} explores the effect of varying treatment intensities on therapeutic outcomes under three treatment strategies. Finally, Section \ref{Sec4} reviews our work and key findings, analyzes the strengths and limitations of our model, and proposes future research directions.

\section{Models and methods}
\label{Sec2}
\subsection{Agent-based model of tumor-immune interactions}
\label{Sec2.1}

Tumor-immune interaction is a complex and precisely regulated process that delineates the dynamic regulatory interplay between the immune system and tumor cells. As the body's defense mechanism, the immune system recognizes and eliminates tumor cells, thereby effectively suppressing tumor growth and dissemination \cite{Dunn.NatRevImmunol.2006,Gubin.ClinCancerRes.2022}. Conversely, tumor cells evade immune surveillance through reduced immunogenicity, secretion of immunosuppressive factors, and upregulation of immune checkpoint expression \cite{Schreiber.Science.2011,Smyth.NatImmunol.2001}. Consequently, deciphering tumor-immune interactions is critical for understanding tumor initiation, progression, and therapeutic interventions.

\begin{figure}[htbp]
	\centering
	\includegraphics[width=14cm]{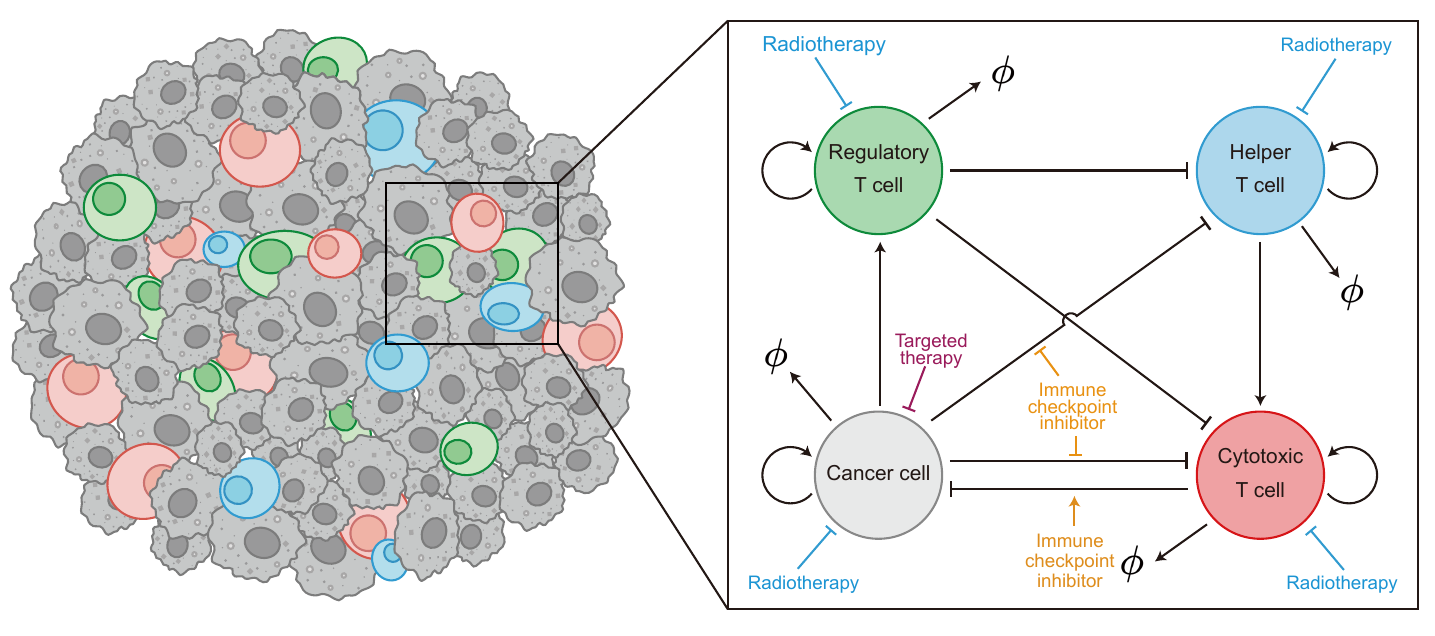}
	\caption{\textbf{Mechanistic diagram of intercellular interactions within the tumor-immune microenvironment.} Tumor cells evade immune surveillance by inducing apoptosis in CTLs and Th cells, while promoting Treg proliferation. Tregs further exert immunosuppressive functions to enhance the death of  CTLs and Th cells. Th cells augment antitumor immune response by stimulating CTL proliferation, with CTLs serving as the key effector cells directly eliminating tumor cells. The model incorporates three treatment mechanisms: radiotherapy (blue), targeted therapy (purple), and immunotherapy (yellow). }
	\label{Fig1}
\end{figure}

In this study, we investigate interactions between tumor cells ($C$) and cytotoxic T cells ($T_c$), helper T cells ($T_h$), and regulatory T cells ($T_r$) (Fig.\ref{Fig1}). To characterize their dynamic behaviors, we assume that each cell undergoes four state transitions: proliferation, death, migration, and stationarity, with these transitions governed by computational rules. Specifically, positive regulation is defined as enhancing the proliferation rate, while negative regulation increases the death rate. Intercellular interactions are primarily influenced by three factors: 
\begin{enumerate}
\item[(1)] Cellular heterogeneity $\mathbf{H}=\{H_C,H_{T_c},H_{T_h},H_{T_r}\}$, which quantifies phenotypic and functional diversity across cell types. 
\item[(2)] Intercellular distance $\xi_{i,j}$ between cell $i$ and $j$, which reflects spatial influences on cellular behavior. 
\item[(3)] Neighborhood cell count $n$ within effective interaction radius $r$, which describes local microenvironmental regulation. 
\end{enumerate}

Additionally, we incorporate tumor drug resistance metrics $\mathbf{D}=\{ x, y\}$ into the model to analyze tumor evolutionary dynamics during treatment and provide a basis for optimizing treatment strategies. By integrating these factors, our computational model can more comprehensively simulate complex intercellular interactions and their dynamic evolution in TME.

\subsubsection{Computational framework for tumor cells}
\label{Sec2.1.1}

Tumor cells usually exhibit high heterogeneity and malignant proliferative potential \cite{Hanahan.Cell.2000,Hanahan.Cell.2011,Hanahan.CancerDiscov.2022}. To quantitatively characterize the heterogeneity among tumor cells, we introduce a metric $H_C \in [0,1]$ that denotes their immunosuppressive potential, with $H_C$ approaching $1$ indicating stronger immunosuppressive capacity, while $H_C$ near $0$ reflects weaker immunosuppression. This metric provides a quantitative basis for analyzing tumor cell dynamics and tumor-immune interactions in the microenvironment.

\textbf{Proliferation rate.} The proliferation rate of tumor cells is defined as:
\begin{equation}
\label{C_pro}
p_{pro}^C=p_{pro, base}^C,
\end{equation}
where $p_{pro,base}^C$ denotes the baseline proliferation rate of tumor cells. In the model, we postulate that tumor cell proliferation depends exclusively on the cell itself and is not influenced by adjacent cells.

\textbf {Death rate.} Within the TME, CTLs induce tumor cell lysis and apoptosis by releasing cytotoxic substances (e.g., perforin and granzymes) and secreting cytokines (e.g., IFN-$\gamma$) \cite{Raskov.BrJCancer.2021,Philip.NatRevImmunol.2022,Giles.Immunity.2023}. Therefore,  based on the interaction between tumor cells and CTLs in Fig.\ref{Fig1}, the death rate of tumor cells is defined as:
\begin{equation}
\label{C_dea}
p_{dea}^C=p_{dea, base}^C+p_{dea, T_c }^C \times \frac{1}{n} (\sum_{k=1}^n \delta_{k, T_c}\times \frac{1}{\xi_{C,T_c}^k} \times H_{T_c}^k),
\end{equation}
where, $p_{dea, base}^C$ denotes the baseline death rate of tumor cells, $p_{dea, T_c}^C$ is the regulatory coefficient of tumor cell death mediated by CTLs, $\delta_{k, T_c}$ is the Dirac delta function indicating whether the $k$-th neighbor of the tumor cell is a CTL, $\frac{1}{\xi_{C, T_c}^k}$ represents the distance-dependent regulatory coefficient between the tumor cell and the CTL at the $k$-th neighbor position, and $H_{T_c}^k$ denotes the cytotoxic activity level of the CTL at that position. Here, $n$ represents the number of neighboring agent sites of a cell. The effectiveness of CTLs in killing tumor cells varies depending on their proximity and cytotoxicity: CTLs located closer to tumor cells and exhibiting higher cytotoxicity demonstrate enhanced killing effectiveness, whereas those farther away with lower cytotoxicity are less effective.

\textbf {Migration rate.} The migration behavior of tumor cells is significantly influenced by the surrounding microenvironment. When cell density in adjacent spaces is high, the available space around tumor cells decreases, and nutrient supply becomes insufficient, thus promoting outward migration toward more favorable growth conditions. Therefore, the migration rate of tumor cells is defined as:
\begin{equation}\label{C_mig}
	p_{mig}^C=p_{mig, base}^C+p_{mig,space}^C \times \frac{1}{n}(n-\sum_{k=1}^n \sum_{j=1}^m \delta_{k, j}), 
\end{equation}
where $p_{mig, base}^C$ denotes the baseline migration rate of tumor cells, $p_{mig, space}^C$ denotes the spatial-information-mediated regulatory coefficient for tumor cell migration, $n-\sum_{k=1}^n \sum_{j=1}^m \delta_{k, j}$ denotes the number of empty agent sites around the current cell, and $m=4$ corresponds to the four cell types considered in the model.

\subsubsection{Computational framework for cytotoxic T cells}
\label{Sec2.1.2}
As the key immune cells responsible for tumor recognition and elimination, CTLs exhibit functional heterogeneity primarily manifested in their cytotoxic levels \cite{Raskov.BrJCancer.2021,Philip.NatRevImmunol.2022,Giles.Immunity.2023,Vasconcelos.CellRep.2015,Khazen.EMBOJ.2021}. We therefore define a heterogeneity metric $H_{T_c} \in [ 0,1]$ to quantify the cytotoxicity levels of CTLs---that is, their tumoricidal capacity. Higher $H_{T_c}$ values indicate stronger tumoricidal capacity, while lower values reflect diminished capacity. In tumor immunology, $H_{T_c}$ can also represent secretion levels of IFN-$\gamma$, granzymes, and perforin.

\textbf{Proliferation rate.} Within the TME, Th cells enhance CTLs proliferation through IL-2 secretion \cite{Raulet.Nature.1982,Liew.NatRevImmunol.2002,Spolski.NatRevImmunol.2018}. Thus, the proliferation rate of CTLs is defined as:
\begin{equation}
\label{Tc_pro}
p_{pro}^{T_c}=p_{pro, base}^{T_c} + p_{pro, T_h}^{T_c} \times \frac{1}{n} (\sum_{k=1}^n \delta_{k,T_h} \times \frac{1}{\xi_{T_c,T_h}^k} \times H_{T_h}^k),
\end{equation}
where $p_{pro, base}^{T_c}$ denotes the baseline proliferation rate of CTLs, $p_{pro, T_h}^{T_c}$ denotes the Th cell-mediated regulatory coefficient for CTL proliferation, $\delta_{k, T_h}$ denotes whether the $k$-th neighboring position contains a Th cell, $\frac{1}{\xi_{T_c, T_h}^k}$ denotes the distance-dependent functional coefficient between CTL and Th cell at the $k$-th neighbor positions, and $H_{T_h}^k$ quantifies the pro-proliferative capacity of the $k$-th Th cell neighbor.  Higher $H_{T_h}$ values indicate stronger pro-proliferative effects on CTLs, while lower values indicate weaker effects.

\textbf{Death rate.} Tumor cells promote CTL apoptosis by expressing PD-L1 molecules \cite{Okazaki.NatImmunol.2013,Patsoukis.SciAdv.2020}. Additionally, as an immunosuppressive T cell subpopulation, Tregs play a critical role in promoting CTL death \cite{Zou.NatRevImmunol.2006,Facciabene.CancerRes.2012,Togashi.NatRevClinOncol.2019}. Therefore, the death rate of CTLs is described as:
\begin{equation}
\label{Tc_dea}
\begin{aligned}
p_{dea}^{T_c} = p_{dea, base }^{T_c} + p_{dea, C}^{Tc} \times \frac{1}{n}(\sum_{k=1}^n \delta_{k,C}\times \frac{1}{\xi_{T_c,C}^k} \times H_{C}^k) + p_{dea,T_r}^{T_c} \times \frac{1}{n}(\sum_{k=1}^n \delta_{k,T_r}\times \frac{1}{\xi_{T_c,T_r}^k} \times H_{T_r}^k),
\end{aligned}
\end{equation}
where $p_{dea, base}^{T_c}$ denotes the baseline death rate of CTLs, $p_{dea, C}^{T_c}$ denotes the death regulation coefficients mediated by tumor cells, $\delta_{k, C}$ denotes whether the $k$-th neighboring position contains a tumor cell, $\frac{1}{\xi_{T_c, C}^k}$ indicates the distance-related functional coefficients between the CTL and the $k$-th neighboring tumor cell, $H_{C}^k$ denotes the inhibitory levels of the tumor cell at that position. Higher values of $H_{C}^k$ indicate stronger suppression of CTL by tumor cells. Tregs influence CTL mortality through mechanisms similar to those of tumor cells.

\textbf {Migration rate.} The migration rate of CTLs is defined as:
\begin{equation}
\label{Tc_mig}
p_{mig}^{T_c} = p_{mig, base}^{T_c} + p_{mig, space}^{T_c} \times \frac{1}{n}(n-\sum_{k=1}^n \sum_{j=1}^m \delta_{k, j}),
\end{equation}
where $p_{mig,base}^{T_c}$ denotes the baseline migration rate of CTLs, and $p_{mig,space}^{T_c}$ denotes the space-mediated mirgation regulatory coefficient of CTLs.

\subsubsection{Computational framework for helper T cells}
\label{Sec2.1.3}
The heterogeneity of Th cells is primarily characterized by their capacity to secrete IL-2, a cytokine that potently enhances CTL proliferation and activation \cite{Liew.NatRevImmunol.2002,Zhou.Immunity.2009}. We define a heterogeneity metric $H_{T_h} \in [ 0,1]$ to quantify the pro-proliferative capacity of Th cells toward CTLs.

\textbf{Proliferation rate.} The proliferation rate of Th cells is defined as:
\begin{equation}
\label{Th_pro}
p_{pro}^{T_h} = p_{pro, base}^{T_h},
\end{equation}
where $p_{pro,base}^{T_h}$ denotes the baseline proliferation rate of Th cells.

\textbf{Death rate.} Similar to their suppression of CTLs, both tumor cells and Tregs inhibit Th cell proliferation and activation in the TME to evade immune surveillance \cite{Zou.NatRevImmunol.2006,Facciabene.CancerRes.2012,Togashi.NatRevClinOncol.2019}. Therefore, the death rate of Th cells is defined as:
\begin{equation}
\label{Th_dea}
\begin{aligned}
p_{dea}^{T_h} = p_{dea, base}^{T_h} + p_{dea, C}^{T_h} \times \frac{1}{n}(\sum_{k=1}^n \delta_{k,C}\times \frac{1}{\xi_{T_h,C}^k} \times H_{C}^k) + p_{dea, T_r }^{T_h} \times \frac{1}{n}(\sum_{k=1}^n \delta_{k,T_r}\times \frac{1}{\xi_{T_h,T_r}^k} \times H_{T_r}^k),
\end{aligned}
\end{equation}
where $p_{dea, base}^{T_h}$ denotes the baseline death rate of Th cells, $p_{dea, C}^{T_h}$ denotes the death regulatory coefficient mediated by tumor cells, and $\frac{1}{\xi_{T_h, C}^k}$ denotes the distance-dependent functional coefficients between the current Th cell and the the $k$-th neighboring tumor cell. The death regulatory parameters of Tregs on Th cells are defined analogously.

\textbf {Migration rate.} The migration rate of Th cells is defined as:
\begin{equation}
\label{Th_mig}
p_{mig}^{T_h} = p_{mig, base}^{T_h} + p_{mig, space}^{T_h} \times \frac{1}{n}(n-\sum_{k=1}^n \sum_{j=1}^m \delta_{k, j}),
\end{equation}
where $p_{mig, base}^{T_h}$ denotes the baseline migration rate of Th cells, and $p_{mig,space}^{T_h}$ denotes the space-mediated regulatory coefficient for Th cell migration.

\subsubsection{Computational framework for regulatory T cells}
\label{Sec2.1.4}
We introduce a Treg heterogeneity metric $H_{T_r} \in [ 0,1]$ to quantify the immunosuppressive level of Tregs, where a higher $H_{T_r}$ value indicates stronger immunosuppression. In tumor immunology, $H_{T_r}$ can be interpreted as representing the secretion levels of IL-10 and TGF-$\beta$ by Tregs, which are known to play important roles in suppressing the differentiation and proliferation of both Th cells and CTLs \cite{Zou.NatRevImmunol.2006,Facciabene.CancerRes.2012,Togashi.NatRevClinOncol.2019}.

\textbf {Proliferation rate.} In the TME, tumor cells shape the formation of an immunosuppressive microenvironment by recruiting and promoting the proliferation of Tregs \cite{Ozga.Immunity.2021,Nagarsheth.NatRevImmunol.2017,Visser.CancerCell.2023,Joyce.Science.2015}. Therefore, the proliferation rate of Tregs is defined as:
\begin{equation}
\label{Tr_pro}
p_{pro}^{T_r} = p_{pro, base }^{T_r} + p_{pro, C }^{T_r} \times \frac{1}{n}(\sum_{k=1}^n \delta_{k,C} \times \frac{1}{\xi_{T_r,C}^k} \times H_{C}^k), 
\end{equation}
where $p_{pro, base}^{T_r}$ denotes the baseline proliferation rate of Tregs, and $p_{pro, C}^{T_r}$ denotes the tumor cell-mediated proliferation regulatory coefficient of Tregs.

\textbf {Death rate.} The death rate of Tregs is defined as:
\begin{equation}
\label{Tr_dea}
p_{dea}^{T_r} = p_{dea, base}^{T_r},
\end{equation}
where $p_{dea,base}^{T_r}$ denotes the baseline death rate of Tregs.

\textbf {Migration rate.} The migration rate of Tregs is defined as:
\begin{equation}
\label{Tr_mig}
p_{mig}^{T_r}=p_{mig, base}^{T_r} + p_{mig, space}^{T_r} \times \frac{1}{n}(n-\sum_{k=1}^n \sum_{j=1}^m \delta_{k, j}),
\end{equation}
where $p_{mig,base}^{T_r}$ denotes the baseline migration rate of Tregs, and $p_{mig, space}^{T_r}$ denotes the space-mediated migration regulatory coefficient of Tregs.

\subsection{Agent-based modeling under different treatment options}
\label{Sec2.2}

To comparatively analyze treatment effects on immune response dynamics and therapeutic outcomes during tumor evolution, we implemented three treatment regimens: 
\begin{enumerate}
\item[(1)] \textbf{Radiotherapy} preferentially kills rapidly dividing cells with impaired DNA repair capacity \cite{De-Ruysscher.NatRevDisPrimers.2019,Chandra.Lancet.2021}. Because tumor cells generally have a higher rate of proliferation and a weaker ability to repair DNA than normal cells, they are more likely to die after radiotherapy \cite{Begg.NatRevCancer.2011,Barker.NatRevCancer.2015}. 
\item[(2)] \textbf{Targeted therapy} achieves selective tumor cell killing by specifically binding to oncogenic sites on tumor cells while sparing normal cells \cite{Choi.iJMS.2023,Ramos.Science.2014}.
\item[(3)] \textbf{Immunotherapy} eliminates tumor cells through immune checkpoint inhibitors (ICIs). In the TME, PD-L1-expressing tumor cells engage PD-1 receptors on T cells to suppress T cell proliferation and impair antitumor immunity, enabling immune escape. PD-1/PD-L1 inhibitors counteract this immune evasion by blocking the PD-1/PD-L1 signaling pathway, thereby reactivating T cell function and enhancing tumor cell killing \cite{Patsoukis.SciAdv.2020,Lu.JAMAOncol.2019,Sun.Immunity.2018}.
\end{enumerate}

\subsubsection{Radiotherapy}
\label{Sec2.2.1}
We hypothesize that radiotherapy increases the death of all cells, but its effect on tumor cell death is significantly higher than on immune cells, thereby achieving effective tumor killing. After applying radiotherapy, the death rates of various cell types are described as follows:
\begin{equation}
\label{cell_dea_rad}
\begin{aligned}
p_{dea}^C&=p_{dea, base}^C + p_{dea,T_c}^C \times \frac{1}{n}(\sum_{k=1}^n \delta_{k, T_c}\times \frac{1}{\xi_{C,T_c}^k} \times H_{T_c}^k) + p_{dea, rad}^C, \\
p_{dea}^{T_c}&=p_{dea, base }^{T_c} + p_{dea, C}^{T_c} \times \frac{1}{n}(\sum_{k=1}^n \delta_{k, C}\times \frac{1}{\xi_{T_c, C}^k} \times H_{C}^k) \\
&\quad{} + p_{dea,T_r}^{T_c} \times \frac{1}{n}(\sum_{k=1}^n \delta_{k,T_r}\times \frac{1}{\xi_{T_c, T_r}^k} \times H_{T_r}^k) + p_{dea, rad}^{T_c},\\
p_{dea}^{T_h}&=p_{dea, base}^{T_h}+p_{dea, C}^{T_h} \times \frac{1}{n}(\sum_{k=1}^n \delta_{k,C}\times \frac{1}{\xi_{T_h,C}^k} \times H_{C}^k)\\
&\quad{} +p_{dea, T_r }^{T_h} \times \frac{1}{n}(\sum_{k=1}^n \delta_{k,T_r}\times \frac{1}{\xi_{T_h,T_r}^k} \times H_{T_r}^k) + p_{dea, rad}^{T_h},\\
p_{dea}^{T_r}&=p_{dea, base}^{T_r} + p_{dea, rad}^{T_r},
\end{aligned}
\end{equation}
where $p_ {dea, rad}^C$, $p_{dea, rad}^{T_c}$, $p_{dea, rad}^{T_h}$, and $p_{dea, rad}^{T_r}$ respectively denote the killing efficacy of radiotherapy on tumor cells, CTLs, Th cells, and Tregs. When simulating the dynamic evolution of the tumor under radiotherapy, equations Eq.\eqref{C_dea}, \eqref{Tc_dea}, \eqref{Th_dea}, and \eqref{Tr_dea} are replaced by Eq.\eqref{cell_dea_rad}.

\subsubsection{Targeted therapy}
\label{Sec2.2.2}
We hypothesize that targeted drugs significantly increase the death rate of tumor cells without affecting other cells. Meanwhile, tumor cells exhibit heterogeneity in their response to targeted drugs, and treatment-induced tumor cell plasticity may lead to drug resistance \cite{Zhang:2021kg,Sharma:2010aa,Ma:2023gd}. Hence, we introduce a drug resistance metric $x\in [0, 1]$ for tumor cells to characterize the dynamic evolution of drug resistance during targeted therapy and its impact on treatment efficacy. Consequently, the death rate of tumor cells under targeted therapy is defined as:
\begin{equation}
\label{C_dea_targeted}
p_{dea}^C=p_{dea,base}^C+p_{dea, T_c }^C \times \frac{1}{n}(\sum_{k=1}^n \delta_{k, T_c}\times \frac{1}{\xi_{C,T_c}^k} \times H_{T_c}^k)+p_{dea, tar}^C\frac{{x_0}^{n_1}}{{x_0}^{n_1}+{x}^{n_1}},
\end{equation}
where $p_{dea, tar}^C$ denotes the regulatory coefficient for targeted drug-induced tumor cell apoptosis, $x$ denotes the drug resistance metric, $x_0$ is the half-saturation constant of drug resistance, and $n_1$ denotes the Hill coefficient. As resistance increases, therapeutic efficacy gradually diminishes. In simulations, Eq.\eqref{C_dea_targeted} replaces Eq.\eqref{C_dea} to better capture resistance effects.

The tumor cell resistance may undergo random changes due to plasticity. We assume the dynamic update of the resistance metric $x$ follows a Markov process. The value at step $t$, $x^t$, is updated to $x^{t+1}$ based on a transition probability conditioned on $x^t$. To maintain the range $0\leq x\leq 1$, we adopt a conditional beta distribution, 
\begin{equation}
\label{Targeted_beta}
P(x^{t+1} = x | x^{t}) \sim B(x | a, b) = \dfrac{\Gamma(a + b)}{\Gamma(a) \Gamma(b)} x^{a-1} (1-x)^{b-1},
\end{equation}
with shape parameters
\begin{equation}
\label{eqB}
a = x^t\eta_1, b = (1-x^t)\eta_1,
\end{equation}
where $\eta_1$ quantifies plasticity. This choice ensures the conditional expectation and variance are: 
$$
\mathrm{E}(x^{t+1} | x^t) = x^t, \mathrm{Var}(x^{t+1} | x^t) = \frac{1}{1+\eta_1} x^t (1-x^t). 
$$
Initial values of $x$ are randomly sampled from $[0, 1]$ before treatment, and updated as each timestep after treatment. 

\subsubsection{Immunotherapy}
\label{Sec2.2.3}

For immunotherapy, we assume that PD-1/PD-L1 inhibitors function in two ways (Fig.\ref{Fig1}):  directly enhancing the killing capacity of CTLs against tumor cells, and reducing tumor cell-mediated immunosuppression of CTLs and Th cells by blocking the PD-1/PD-L1 pathway. The death rates of tumor cells, CTLs, and Th cells during immunotherapy are defined as:
\begin{equation}
\label{cell_dea_ICI}
\begin{aligned}
p_{dea}^C&=p_{dea, base}^C+p_{dea, T_c }^C\times \frac{1}{n}(\sum_{k=1}^n \delta_{k, T_c}\times \frac{1}{\xi_{C,T_c}^k}\times H_{T_c}^k)\times(p_{dea,ICI}^C \frac{{y_0}^{n_2}}{{y_0}^{n_2}+y^{n_2}}),\\
p_{dea}^{T_c}&=p_{dea, base }^{T_c}+p_{dea, C}^{T_c}\times \frac{1}{n}(\sum_{k=1}^n \delta_{k,C}\times \frac{1}{\xi_{T_c,C}^k}\times H_{C}^k )\times(p_{dea,ICI}^{T_c}\frac{{y_k}^{n_2}}{{y_k}^{n_2}+{y_0}^{n_2}})\\
&\quad{}+p_{dea,T_r}^{T_c}\times\frac{1}{n}(\sum_{k=1}^n \delta_{k,T_r}\times\frac{1}{\xi_{T_c,T_r}^k}\times H_{T_r}^k),\\
p_{dea}^{T_h}&=p_{dea, base}^{T_h}+p_{dea, C}^{T_h} \times \frac{1}{n} (\sum_{k=1}^n \delta_{k,C}\times \frac{1}{\xi_{T_h,C}^k} \times H_{C}^k) \times( p_{dea, ICI}^{T_h} \frac{{y_k}^{n_2}}{{y_k}^{n_2}+{y_0}^{n_2}})\\ 
&\quad{}+p_{dea, T_r }^{T_h} \times \frac{1}{n}(\sum_{k=1}^n \delta_{k,T_r}\times \frac{1}{\xi_{T_h,T_r}^k} \times H_{T_r}^k),
\end{aligned}
\end{equation}
where $p_{dea, ICI}^C$, $p_{dea, ICI}^{T_c}$, and $p_{dea, ICI}^{T_h}$ denote the regulatory coefficients of ICI treatment on tumor cells, CTLs, and Th cells, respectively. Here, $y$ is the ICI resistance metric for tumor cells, and $y_k$ is the resistance level of the $k$-th tumor cell adjacent to CTLs or Th cells. The parameter $y_0$ denotes the half-saturation constant, and $n_2$ is the Hill coefficient. In simulations, Eq.\eqref{cell_dea_ICI} replaces Eqs.\eqref{C_dea}, \eqref{Tc_dea}, and \eqref{Th_dea} to represent the effect of immunotherapy.

As with targeted therapy, we model resistance dynamics using a conditional beta distribution:
\begin{equation}
\label{ICI_beta}
P(y^{t+1} = y | y^t) \sim B(y | a,b)
\end{equation}
with  shape parameters
$$a=y^t\eta_1, b=(1-y^t)\eta_2.$$ 
Thus, the conditional expectation and variance of $y^{t+1}$ are:
$$
E(y^{t+1} | y^t) = y^t, \mathrm{Var}(y^{t+1} | y^t) = \frac{1}{1 + \eta_2} y^t (1-y^t).
$$
Initial values $y$ are randomly sampled from $[0, 1]$ and updated at each timestep following treatment.

\subsection{Agent-based discrete stochastic simulation}
\label{Sec2.3}

\subsubsection{Simulation workflow}
\label{Sec2.3.1}
We develop an agent-based discrete stochastic simulation scheme to model the spatiotemporal dynamics of tumor-immune system interactions and to evaluate the effects of different treatment strategies. The model employs a discretized spatiotemporal framework to accurately characterize the dynamic interactions of key cellular components in the TME. These cells follow specific behavioral rules to perform fundamental biological activities in a two-dimensional discrete space, such as proliferation, death, and migration (Fig.\ref{Fig1}).

The model uses a rectangular grid for spatial discretization. Although cell sizes exhibit heterogeneity in real biological systems, we simplify the model by assigning each cell to a single grid unit. This discretization enables local interactions between cells and their Moore neighborhood, balancing computational efficiency with spatial cell-cell interactions \cite{West.TRENDSCELLBIOL.2023}.

\begin{figure}[htbp]
\centering
\includegraphics[width=13cm]{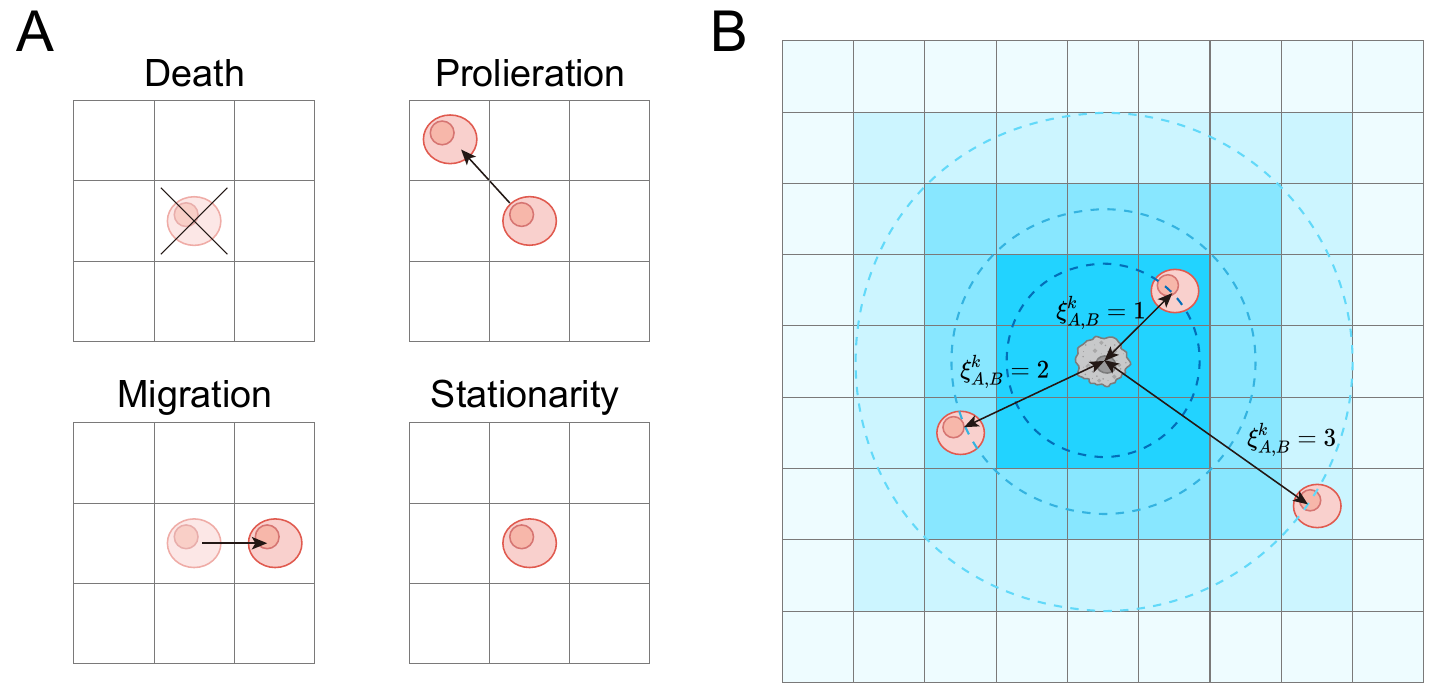}
\caption{\textbf{Schematic diagram of cell fate decision rules.} (A) Rules for updating cellular behaviors: death, proliferation, migration, and stationarity. (B) Definition of distance from a cancer cell. }
\label{Fig2}
\end{figure}

Stochastic simulation is used to simulate tumor progression dynamics. During each time step $\Delta t$, each cell makes a fate decision based on the transition rates defined above. For a transition rate $p$ (as calculated in Sections \ref{Sec2.1} and \ref{Sec2.2}), the probability of a cell-fate transition within $\Delta t$ is $p\times \Delta t$. When a cell chooses death, proliferation, migration, or otherwise remains stationary, the corresponding update is applied (Fig.\ref{Fig2}A):
\begin{description}
\item[Death:] The cell is removed from the system, and the agent unit becomes empty.
\item[Proliferation:] The original cell remains, and a randomly selected empty unit in the Moore neighborhood (eight adjacent units) is populated with a daughter cell. The daughter cell inherits the type and heterogeneity metrics of the mother cell.
\item[Migration:] The cell is moved to a randomly selected adjacent empty unit. 
\item[Stationarity:] The cell remains in place with no change in state.
\end{description}

When computing cell proliferation and death rates, the model incorporates influences from other cells within three concentric layers centered on the target cell (Fig.\ref{Fig2}B). The model uses a distance-dependent regulatory mechanism. For example, in tumor cell-CTL interactions, $\xi^k_{C, T_C}$ represents the distance between a tumor cell and a CTL, and $\frac{1}{\xi^k_{C, T_C}}$ is the associated distance-weighted regulatory coefficient. This factor is used in Eq.\eqref{C_dea} to quantify CTL-mediated tumor killing. This mechanism captures the distance-decay effect, where CTL killing efficacy declines with increasing distance, reflecting physiological behavior and enabling realistic modeling of immune evasion. 

\begin{figure}[htbp] 
	\centering
	\includegraphics[width=16cm]{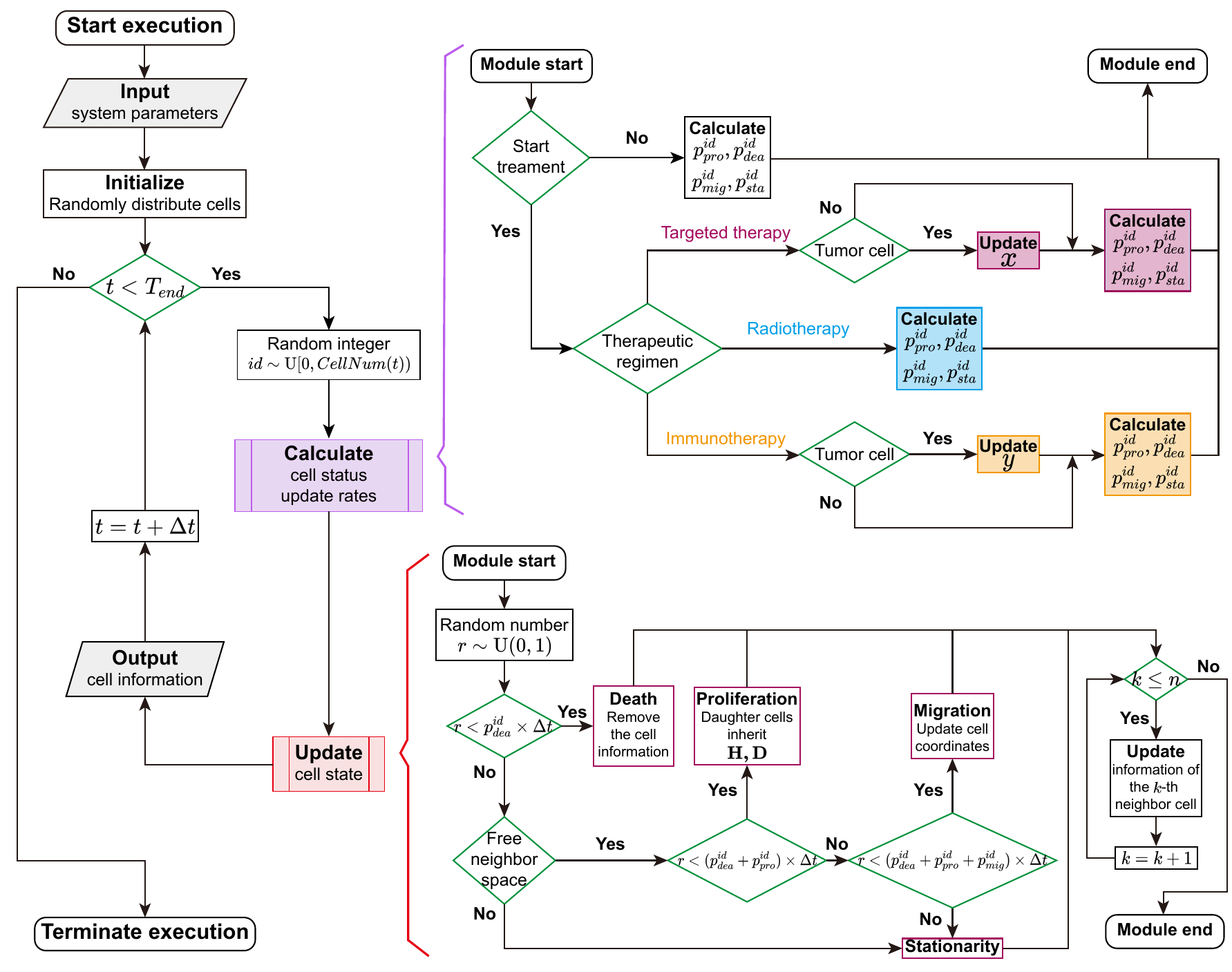}
	\caption{\textbf{Stochastic simulation computational workflow.} The left panel shows the main procedural framework. The two key modules used for calculating cell update rates (purple) and updating states (red) are shown as color-coded flowcharts.}
	\label{Fig3}
\end{figure}

The model is implemented in the \verb|C++| programming language and runs on both Linux and Microsoft Windows systems. The program workflow is shown in Fig.\ref{Fig3} and Algorithm \ref{alg:abm}. The main steps are:
\begin{enumerate}
\item[(1)] System initialization: Read all parameter configuration files and set up the simulation environment.
\item[(2)] Cell placement: Tumor cells are positioned in the center of the space; immune cells are randomly placed around the tumor to simulate a solid tumor environment.
\item[(3)] Cell state update: At each time step $\Delta t$, one cell is randomly selected and updated according to behavior rules and transition rates.
\item[(4)] Data recording: Key outputs---including spatial positions and neighborbood relationship---are recoded in real time. 
\end{enumerate}

\begin{algorithm}
	\caption{Agent based modeling} 
	\label{alg:abm} 
	\begin{algorithmic}[1]
		\REQUIRE System parameters and cell state transition parameters 
		\ENSURE Cell information matrix and population counts per type
		
		\STATE \textit{Initialize:} randomly distributed cells
		\FOR{$t = 0$ \TO $T_{end}$}
		\STATE Select a random cell: $id \gets \lfloor \text{rand}() \times CellNum \rfloor$
		\STATE \textit{Calculate transition rates:}
		\IF{No therapy}
		\STATE Calculate: $p^{id}_{dea}, p^{id}_{pro}, p^{id}_{mig}$ according to Eq.\eqref{C_pro}-\eqref{Tr_mig}
		\ELSIF{Radiotherapy}
		\STATE Calculate: $p^{id}_{dea}, p^{id}_{pro}, p^{id}_{mig}$ according to Eq.\eqref{cell_dea_rad} 
		\ELSIF{Targeted therapy}
		\IF{$T_c$}
		\STATE Update: $x^t$ according to Eq.\eqref{Targeted_beta}
		\ENDIF
		\STATE Calculate: $p^{id}_{dea}, p^{id}_{pro}, p^{id}_{mig}$ according to Eq.\eqref{C_dea_targeted} 
		\ELSE
		\IF{$T_c$}
		\STATE Update: $y^t$ according to Eq.\eqref{ICI_beta}
		\ENDIF
		\STATE Calculate: $p^{id}_{dea}, p^{id}_{pro}, p^{id}_{mig}$ according to Eq.\eqref{cell_dea_ICI}
		\ENDIF
		\STATE \textit{Cell-fate decision:}
		\STATE Generate a random number: $r \gets \text{rand}(0,1)$								
		\IF{$r < p^{id}_{dea} \times \Delta t$}  
		\STATE Death: remove the cell
		\STATE $CellNum \gets CellNum - 1$
		\ELSIF{free neighboring space}				
		\IF{$r < (p^{id}_{dea} + p^{id}_{pro}) \times \Delta t$}
		\STATE Proliferation: daughter cells inherits $H_{id}, x, y$ 
		\STATE $CellNum \gets CellNum + 1$
		\ELSIF{$r < (p^{id}_{dea} + p^{id}_{pro} + p^{id}_{mig}) \times \Delta t$}
		\STATE Migration: update coordinates
		\ELSE
		\STATE Stationarity: no change in cell state
		\ENDIF
		\ELSE
		\STATE Stationarity: no change in cell state
		\ENDIF 
		\FOR{$k = 0$ \TO $n$}
		\STATE Update: information for the $k$-th neighbor cell 
		\ENDFOR
		\ENDFOR
	\end{algorithmic} 
\end{algorithm}

\subsubsection{Model parameters}
\label{Sec2.3.2} 
All key parameters used in the model simulation and listed in Table \ref{table:par}. The total simulation time $T_{end}$, time step $\Delta t$, spatial dimensions, and initial cell counts are kept fixed as basic parameters. To realistically reflect the clinical tumor microenvironment, the initial fractions of different cell types are chosen within ranges based on the cellular composition observed in cancer patients. Initial cell numbers are randomly sampled from these ranges. Other parameters, including cell transition rates and regulation coefficients, are configured based on specific treatment scenarios.

\begin{table}[htbp]
\centering
\caption{Agent-based model parameters value}
\label{table:par}
\begin{tabular}{l l l l}
\hline
	Parameter & Value & Unit$^*$ & Description \\  
\hline
	SpaceWidth & $100$ & - & Width of the two-dimensional space  \\
	SpaceLength & $100$ & - & Length of the two-dimensional space  \\
 	$T_{end}$ & $60$ & day & Total simulation time\\
 	$\Delta t$ & $0.001$ & day & Simulation time step \\
  	$N_{C, ini}$ & $[580,620]$ & - & Initial number of tumor cells  \\
  	$N_{T_c, ini}$ & $[160,180]$ & - & Initial number of cytotoxic T cells \\
  	$N_{T_r, ini}$ & $[40,60]$ & - & Initial number of regulatory T cells   \\
	$N_{T_h, ini}$ & $[190,210]$ & - & Initial number of helper T cells   \\
	$n$ & $48$ & - & Number of neighboring agents (3 concentric layers) \\
	$p_{dea, base}^C$ & $10.00$ & $1/\mathrm{day}$ & Baseline death rate of tumor cells ($C$) \\
	$p_{pro, base}^C$ & $250.00$ & $1/\mathrm{day}$ & Baseline proliferation rate of $C$ \\
	$p_{mig, base}^C$ & $100.00$ & $1/\mathrm{day}$ & Baseline migration rate of $C$ \\
	$p_{mig, space}^C$ &$100.00$ & $1/\mathrm{day}$ & Spatially mediated regulatory coefficient of $C$ migration\\
     $p_{dea,T_c}^C$ & $10500.00$ & $1/\mathrm{day}$ & $T_c$-mediated regulation coefficient of $C$ death\\
	$p_{dea, rad}^C$ & $200.00$ & $1/\mathrm{day}$ & Radiotherapy-induced death rate increase for $C$\\
	$p_{dea, tar}^C$ & $600.00$ & $1/\mathrm{day}$ & Targeted therapy-induced death rate increase for $C$\\
	$x_0$ & $0.5$ & - & Half-saturation constant for targeted therapy resistance \\
     $n_1$ & $2$ & - & Hill coefficient for targeted therapy resistance \\
	$\eta_1$ & $5000$ & - & Resistance update parameter (targeted therapy) \\
	$p_{dea, ICI}^C$ & $20000.00$ & $1/\mathrm{day}$ & Immunotherapy-induced death rate increase for $C$\\
	$y_0$ & $0.5$ & - & Half-saturation constant for ICI resistance \\
     $n_2$ & $6$ & - & Hill coefficient for ICI resistance \\
	$\eta_2$ & $100$ & - & Resistance update parameter (immunotherapy) \\
	$p_{dea, base}^{T_c}$ & $30.00$ & $1/\mathrm{day}$ & Baseline death rate of CTLs ($T_c$) \\
     $p_{pro, base}^{T_c}$ & $150.00$ & $1/\mathrm{day}$ & Baseline proliferation rate of $T_c$ \\
	$p_{mig, base}^{T_c}$ & $600.00$ & $1/\mathrm{day}$ & Baseline migration rate of $T_c$ \\
     $p_{mig, space}^{T_c}$ & $600.00$ & $1/\mathrm{day}$ & Spatially mediated regulatory coefficient of $T_c$ migration \\
     $p_{dea, C}^{T_c}$ & $10500.00$ & $1/\mathrm{day}$ & $C$-mediated regulation coefficient of $T_c$ death \\
     $p_{dea,T_r}^{T_c}$ & $8700.00$ & $1/\mathrm{day}$ & $T_r$-mediated regulation coefficient of $T_c$ death\\
    $p_{pro,T_h}^{T_c}$ & $10600.00$ & $1/\mathrm{day}$ & $T_h$-mediated regulation coefficient of $T_c$ death \\
    $p_{dea, rad}^{T_c}$ & $150.00$ & $1/\mathrm{day}$ & Radiotherapy-induced death rate increase for $T_c$ \\
    $p_{dea, ICI}^{T_c}$ & $50.00$ & $1/\mathrm{day}$ & Immunotherapy-induced death rate increase for $T_c$ \\
    $p_{dea, base}^{T_h}$ & $30.00$ & $1/\mathrm{day}$ & Baseline death rate of Th cells ($T_h$) \\
    $p_{pro, base}^{T_h}$ & $150.00$ & $1/\mathrm{day}$ & Baseline proliferation rate of $T_h$ \\
    $p_{mig, base}^{T_h}$ & $550.00$ & $1/\mathrm{day}$ & Baseline migration rate of $T_h$ \\
    $p_{mig, space}^{T_h}$ & $600.00$ & $1/\mathrm{day}$ & Spatial mediated regulatory coefficient of $T_h$ migration \\
    $p_{dea, C}^{T_h}$ & $8900.00$ & $1/\mathrm{day}$ & $C$-mediated regulation coefficient of $T_h$ death \\
    $p_{dea, T_r}^{T_h}$ & $8700.00$ & $1/\mathrm{day}$ & $T_r$-mediated regulation coefficient of $T_h$ death \\
    $p_{dea, Rad}^{T_h}$ & $150.00$ & $1/\mathrm{day}$ & Radiotherapy-induced death rate increase for $T_h$ \\
    $p_{dea, ICI}^{T_h}$ & $50.00$ & $1/\mathrm{day}$ & Immunotherapy-induced death rate increase for $T_h$ \\
    $p_{dea, base}^{T_r}$ & $30.00$ & $1/\mathrm{day}$ & Baseline death rate of Treg cells ($T_r$) \\
    $p_{pro, base}^{T_r}$ & $50.00$ & $1/\mathrm{day}$ & Baseline proliferation rate of $T_r$ \\
    $p_{mig, base}^{T_r}$ & $700.00$ & $1/\mathrm{day}$ & Baseline migration rate of $T_r$ \\
    $p_{mig, space}^{T_r}$ & $600.00$ & $1/\mathrm{day}$ & Spatial mediated regulatory coefficient of $T_r$ migration \\
    $p_{pro, C}^{T_r}$ & $8700.00$ & $1/\mathrm{day}$ & $C$-mediated regulation coefficient of $T_r$ proliferation \\
    $p_{dea, Rad}^{T_r}$ & $200.00$ & $1/\mathrm{day}$ & Radiotherapy-induced death rate increase for $T_r$\\  
\hline
\end{tabular}
\begin{minipage}{15cm}
\begin{enumerate}
\item[$^*$] The state transition rates correspond to the rate of a single cell in each cell fate decision process. Since we select only one cell in each step and have a total of $100\times 100 = 10^4$ agent units, each cell has a probability of $1/10^4$ of being selected in one time step. Consequently, the actual unit of the transition rates should be $1/(10^4\times \mathrm{day})$.
\end{enumerate}
\end{minipage}
\end{table}

\section{Results}
\label{Sec3}

\subsection{Stochastic simulation of tumor-immune interactions}
\label{Sec3.1}

\begin{figure}[htbp]
	\centering
	\includegraphics[width=14cm]{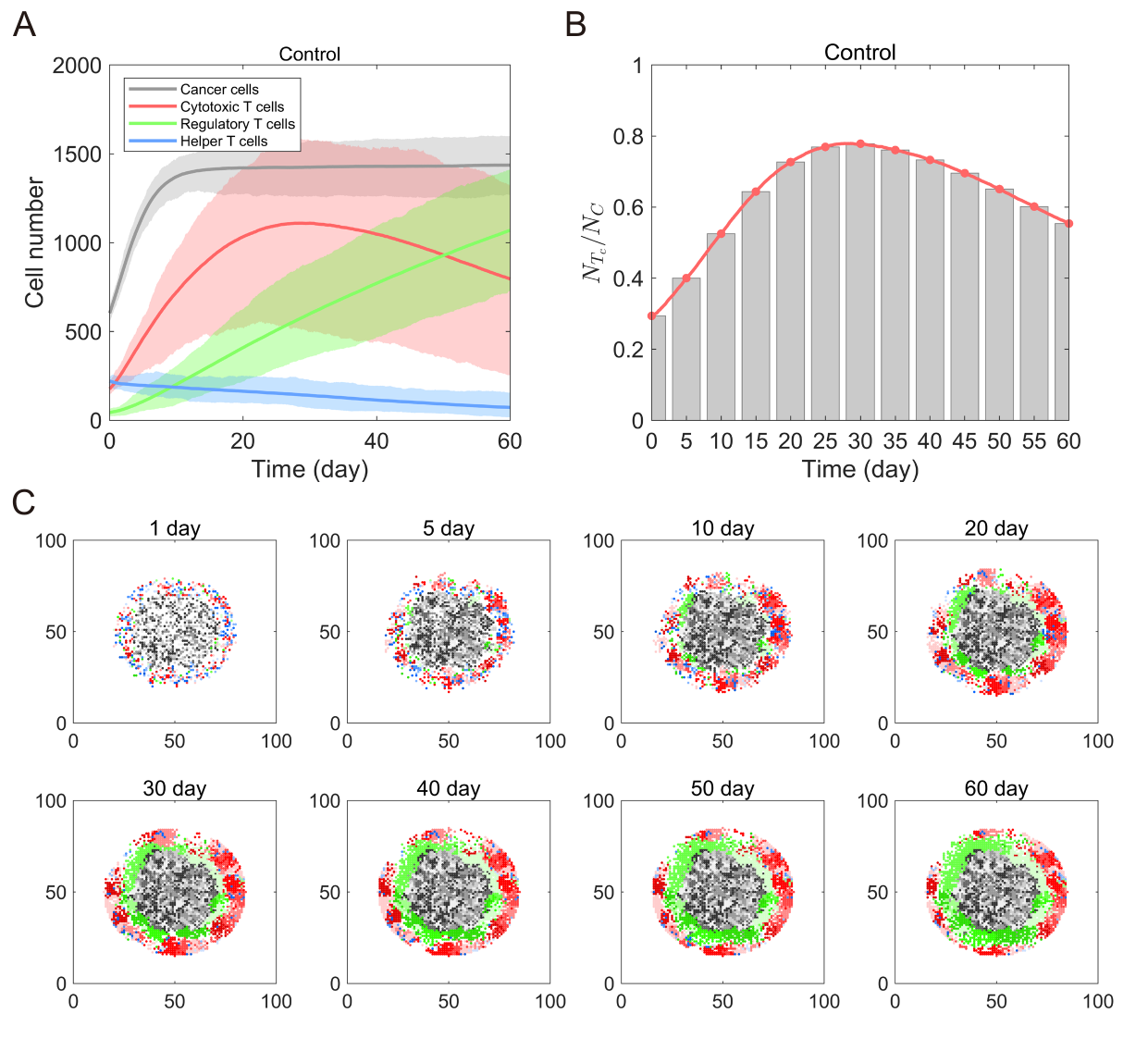}
	\caption{\textbf{Simulation of tumor evolution under tumor-immune cell interactions.} (A) Cell population dynamics over $60$ days, with shaded regions showing the ranges of $100$ stochastic simulations and solid lines indicating their mean values. (B) The ratio of CTLs to tumor cells over time. (C) Two-dimensional spatial distributions recorded every $5$ days: gray dots for tumor cells, red dots for CTLs, green dots for Tregs, blue dots for Th cells. Color intensity represents heterogeneity levels of various cell types.}
	\label{Fig4}
\end{figure}

To investigate the impact of the immune system on tumor growth, we first simulate the spatiotemporal dynamics of tumor growth under untreated conditions based on Eqs.\eqref{C_pro}-\eqref{Tr_mig}. To ensure the reliability of the simulation results, we perform $100$ independent stochastic simulations. Fig.\ref{Fig4} shows the temporal trends of different cell populations across the $100$ simulations, along with spatial distribution characteristics at different time points.

The simulation results show that tumor cell numbers increase rapidly at time, then gradually stabilize (Fig.\ref{Fig4}A). This trend may be attributed to sparse initial cell distribution, abundant nutrient supply, and limited immune cell infiltration during early tumor development, which collectively create favorable conditions for rapid proliferation. Concurrently, CTL numbers rise significantly during the first $20$ days before gradually declining. Treg populations continuously increase in the TME, while Th cell numbers slowly decrease (Fig.\ref{Fig4}A). The ratio of CTLs to tumor cells, which reflects immune killing capacity in the TME, exhibits a dynamic trend of initial increase followed by decline (Fig.\ref{Fig4}B).

These results demonstrate complex interactions between immune cell dynamics and tumor growth in the TME. The initial increase in CTLs reflects the immune system's early response to tumor presence, while the subsequent decline reveals immunosuppressive effects and spatial competition triggered by tumor cells and Tregs. The continuous rise in Tregs further exacerbates the formation of an immunosuppressive microenvironment, limiting the function of CTLs and Th cells. The gradual decrease in Th cells can be attributed to immunosuppression induced by both tumor cells and Tregs. The changes in immune killing capacity in the microenvironment illustrate spontaneous evolution toward an immunosuppressive state in the absence of therapeutic intervention. Overall, these dynamic changes revealed key mechanisms of tumor immune evasion and provide important insights into the complexity of tumor-immune interactions.

Additionally, from a spatial distribution perspective, immune cells are uniformly distributed at the tumor periphery during the first $5$ days (Fig.\ref{Fig4}C). Subsequently, from day $10$ to $20$, immune cells gradually migrated toward the tumor region, forming an ``immune infiltration'' pattern. Between days $30$ and $60$, Tregs form a ring-like ``barrier'' around the tumor periphery, physically separating tumor cells from CTLs and thereby attenuating CTL tumoricidal effects (Fig.\ref{Fig4}C). In tumor immunology, this phenomenon is termed ``immune privilege'' \cite{Joyce.Science.2015}, representing one of the key mechanisms of tumor immune evasion. However, the Treg encirclement also partially constrains further tumor expansion.

\subsection{Tumor progression under different treatment methods}
\label{Sec3.2}
\subsubsection{Radiotherapy}
\label{Sec3.2.1}

\begin{figure}[htbp]
	\centering
	\includegraphics[width=14cm]{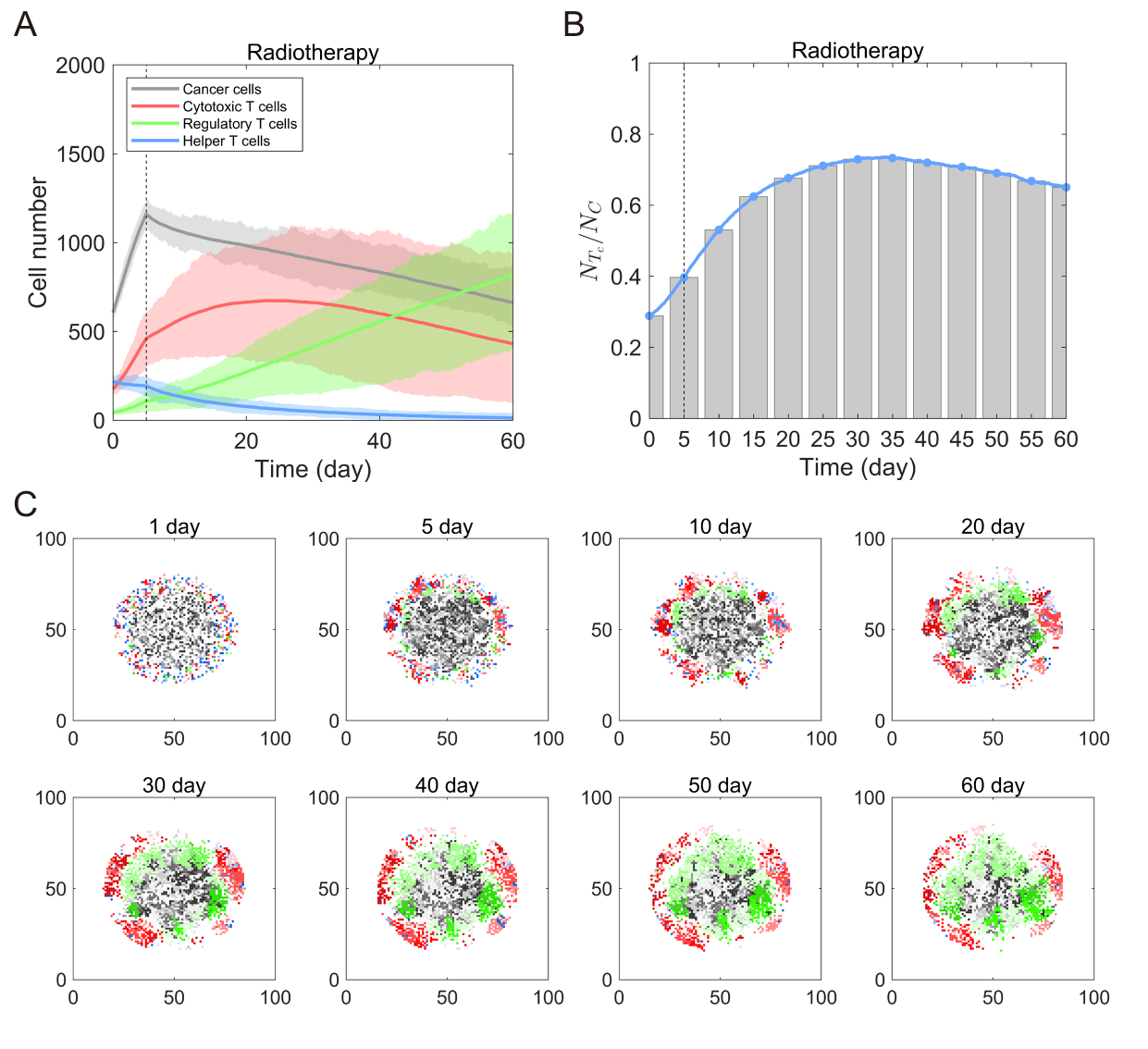}
	\caption{\textbf{Simulated tumor-immune system dynamics under radiotherapy.} (A) Radiotherapy is administered from day $5$ to day $60$ on top of baseline simulation. Solid lines show averaged cell population changes across $100$ stochastic simulations. (B) Change in the ratio of CTLs to the number of tumor cells over time under radiotherapy. (C) Tumor-immune system spatial distributions at different time points.}
	\label{Fig5}
\end{figure}

To elucidate tumor-immune system dynamics and therapeutic effects during radiotherapy, we apply Eq.\eqref{cell_dea_rad} for the cell death rates. Simulation results show that after radiotherapy is initiated on day $5$, tumor growth is significantly suppressed, with rapid tumor cell depletion. By day $60$, tumor cell counts decrease by approximately $50\%$ compared to baseline simulations (Fig.\ref{Fig5}A). During radiotherapy, all immune cell populations are also further reduced relative to baseline. After treatment initiation, the killing capacity of T cells increases rapidly, then declines slightly, and finally stabilizes at around $0.65$ (Fig.\ref{Fig5}B). Comparison with baseline simulations reveals that although radiotherapy damages immune cells to some extent, it ultimately enhances immune-killing capacity and improves the TME in the later stages of the simulation.

From a spatial perspective, starting from day $30$, the effects of radiotherapy became evident, manifested by halted tumor volume expansion followed by gradual shrinkage, particularly with significantly reduced cell density in central regions and loosened tumor tissue (Fig.\ref{Fig5}C). Meanwhile, Tregs remain distributed at the tumor periphery, but their encirclement becomes less compact, providing increased opportunities for CTL infiltration. With continued radiotherapy, both tumor cells and Tregs display progressively lighter coloration, indicating substantial phenotypic alterations. This observation suggests that radiotherapy, through selective pressure, may have preferentially preserved subpopulations with weaker immunosuppressive capacity, thereby modifying the immunosuppressive properties of the TME.

\subsubsection{Targeted therapy}
\label{Sec3.2.2}

To investigate the dynamic changes in the tumor-immune system under targeted therapy, we employ Eq.\eqref{C_dea_targeted} to describe the tumor cell death rate. Under targeted drug therapy, the tumor cell death rate significantly increases, and the therapeutic efficacy is simultaneously modulated by the drug resistance.

\begin{figure}[htbp]
\centering
\includegraphics[width=14cm]{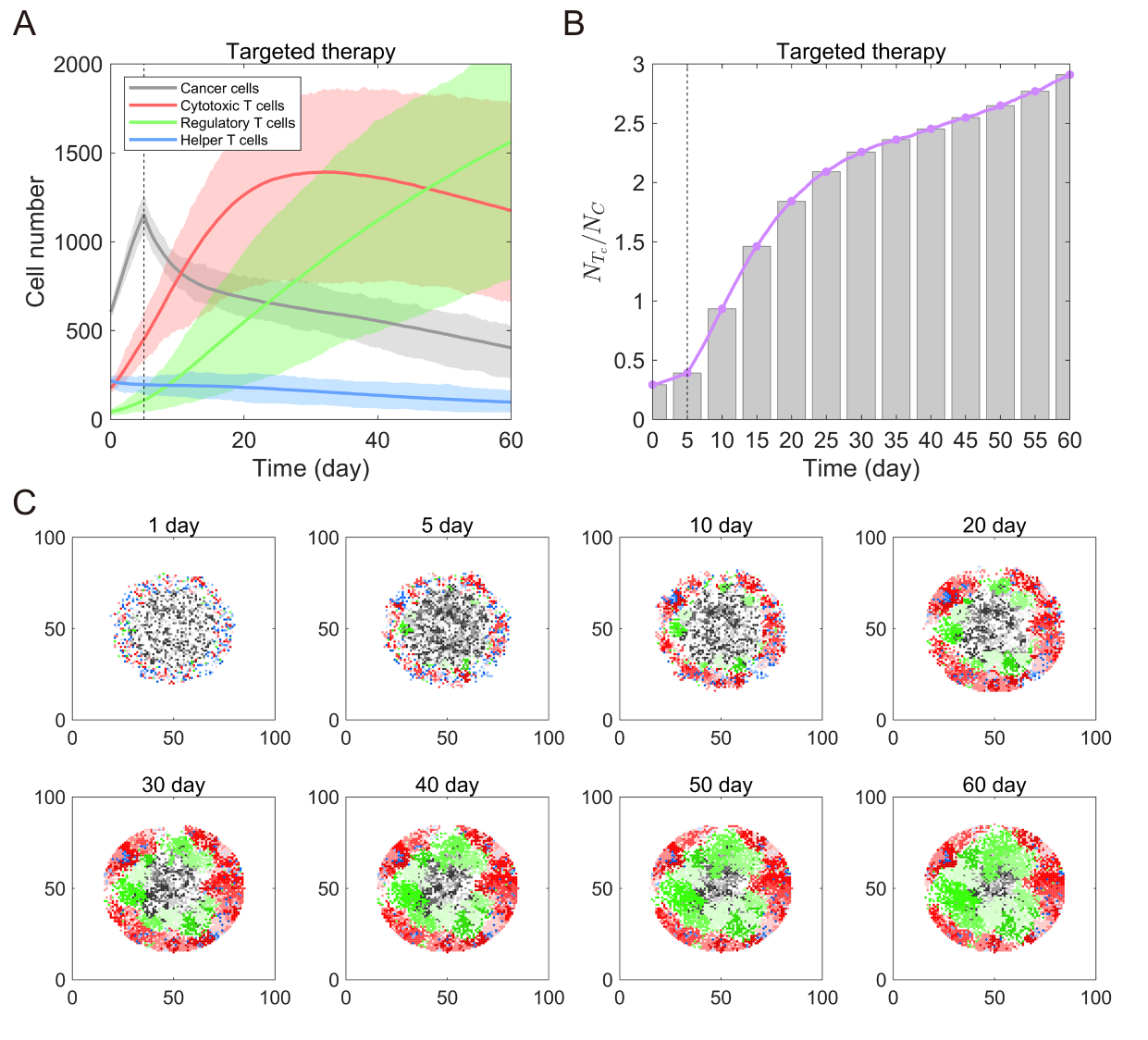}
\caption{\textbf{Simulation of tumor-immune system dynamics under targeted therapy.} (A) Cell population changes from day $5$ to day $60$ with targeted therapy on top of baseline simulation. Solid lines show mean values from $100$ stochastic simulations (B) CTL to tumor cell number ratio changes over time. (C) Two-dimensional spatial distributions of the tumor-immune system at selected timepoints.}
	\label{Fig6}
\end{figure}

Model simulations show that after targeted therapy initiation, the tumor cell population decreases rapidly, reaching approximately $60\%$ of its initial value by the end of the simulation, indicating significant anti-tumor efficacy (Fig.\ref{Fig6}A). Compared to the untreated group, targeted therapy leads to a notable increase in CTLs and Tregs, with a slight rise in Th cells (Fig.\ref{Fig6}A). The increased immune cell populations may be attributed to two factors: (1) targeted drugs specifically kill tumor cells without directly affecting immune cells, and (2) massive tumor cell death provides more space and resources for immune cell expansion.

Following treatment initiation, the immune-killing capacity in the TME shows significantly accelerated growth (Fig.\ref{Fig6}B). Up to day $30$, the rate of increase slows due to reduced CTL numbers. Nevertheless, the immune-killing capacity maintains an upward trend, ultimately reaching a peak level of $2.91$ at the end of the simulation. The spatial distribution of the tumor-immune system reveals that from day $20$ onward, tumor volume decreases significantly, and tumor cells become more sparsely distributed (Fig.\ref{Fig6}C). These results demonstrate that targeted therapy not only inhibits tumor growth but also enhances immune activity by remodeling the TME, offering new perspectives for cancer treatment.

\subsubsection{Immunotherapy}
\label{Sec3.2.3}

Immune checkpoint inhibitors exert their therapeutic effects by enhancing the tumor-killing capacity of CTLs and suppressing tumor immune evasion. To quantitatively analyze the tumor-immune system dynamics under immunotherapy, we integrate the death rates of tumor cells, CTLs, and Th cells using Eq.\eqref{cell_dea_ICI} to simulate the therapeutic effects of PD-1/PD-L1 inhibitors.

\begin{figure}[htbp]
	\centering
	\includegraphics[width=14cm]{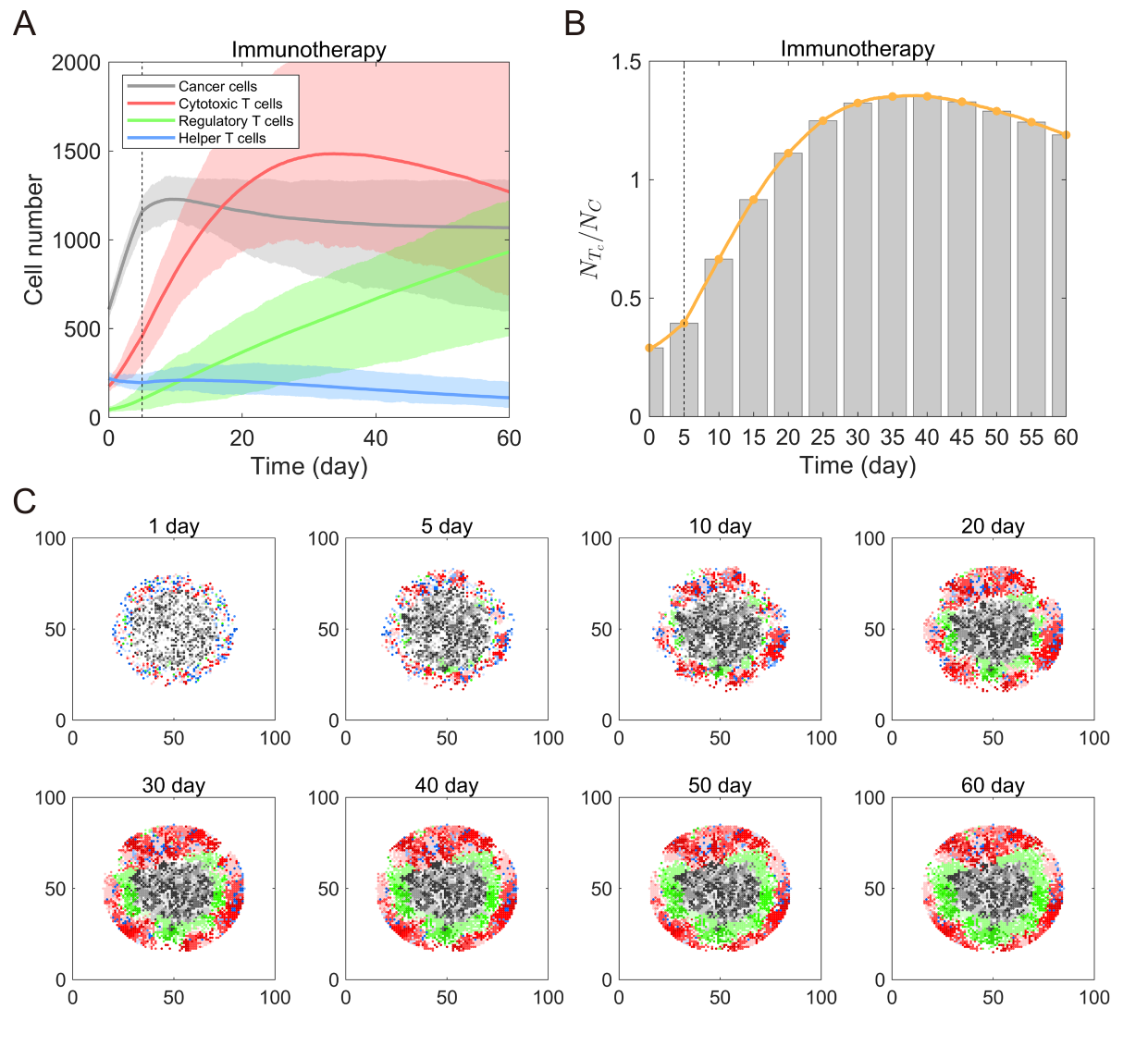}
	\caption{\textbf{Simulated tumor-immune system dynamics under immunotherapy.} (A) Cell population changes from day $5$, when immune checkpoint inhibitor treatment begin, until day $60$, when the simulation ended. Solid lines show averages of $100$ simulations. (B) Changes in the ratio of CTLs to tumor cells over time under immunotherapy. (C) Two-dimensional spatial distributions of the tumor-immune system at selected time points.} 
	\label{Fig7}
\end{figure}

Stochastic simulation results demonstrate that immunotherapy effectively suppresses tumor growth, as evidenced by a gradual decline in tumor cell numbers (Fig.\ref{Fig7}A). CTL counts increase rapidly during the initial treatment phase, then decrease slightly, yet ultimately remain higher than tumor cell counts. Although Treg numbers show slow but steady growth, they remain consistently lower than CTL counts. Th cell numbers stabilize after a transient increase. The CTL to tumor cell ratio shows significant improvement, peaking at $1.35$ around day $38$ and maintaining high levels until treatment termination (Fig.\ref{Fig7}B). These results indicate that immune checkpoint inhibitors not only effectively control tumor progression but also enhance the proliferation of CTLs with potent tumor-killing functions.

Spatially, tumor volume stabilizes after approximately $15$ days of treatment, halting further expansion (Fig.\ref{Fig7}C). This phenomenon may result from immune checkpoint inhibitors restoring T cell function and enhancing anti-tumor activity via PD-1/PD-L1 blockade. By mid to late stages of the simulation, most CTLs appear dark red, while Th cells display dark blue coloration (Fig.\ref{Fig7}C). This indicates that immunotherapy not only reduces the mortality of immune cells but also enhances functional diversity through increased heterogeneity. Specifically, dark red CTLs represent subpopulations with high cytotoxicity, while dark blue Th cells denote subpopulations with elevated IL-2 secretion capacity. This enhanced heterogeneity further optimizes immune responses within the TME, thereby strengthening anti-tumor immunity.

\subsection{Dynamics of drug-resistance under different treatment options}
\label{Sec3.3}

To thoroughly analyze the dynamic changes in tumor cell drug resistance under targeted therapy and immunotherapy, we introduce two drug resistance metrics: targeted therapy resistance $x$ and immunotherapy resistance $y$. During simulations of targeted therapy and immunotherapy, these resistance metrics are updated according to Eq.\eqref{Targeted_beta} and \eqref{ICI_beta}, respectively. These metrics play crucial regulatory roles in the tumor cell death rate. 

\begin{figure}[htbp]
	\centering
	\includegraphics[width=14cm]{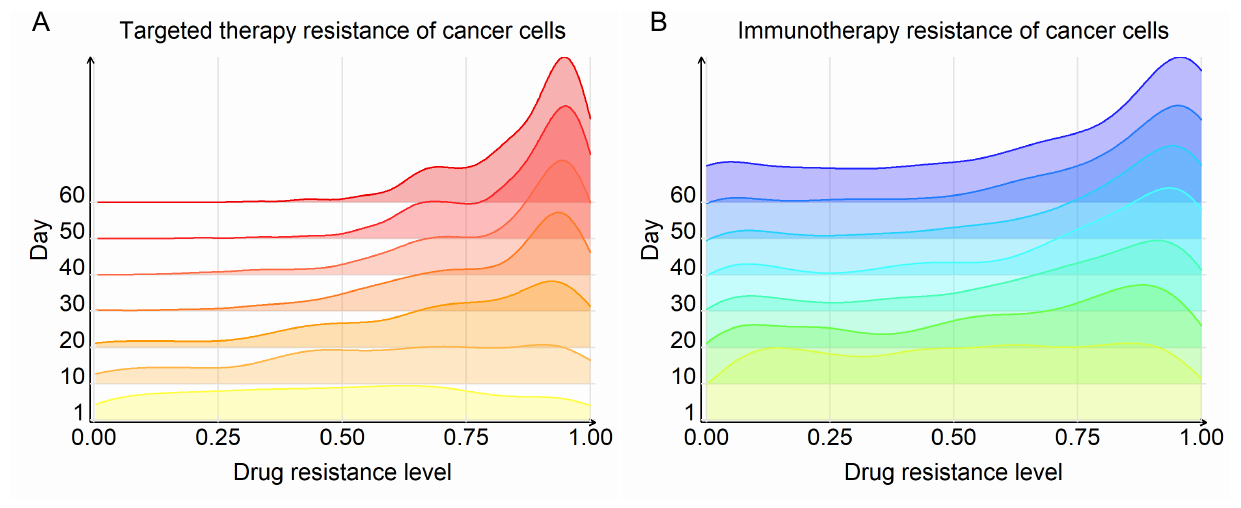}
	\caption{\textbf{Dynamic changes of tumor cell drug resistance during targeted therapy and immunotherapy.} The probability distribution of drug resistance metrics across all tumor cells is calculated every $10$ days. (A) Distribution under targeted therapy. (B) Distribution under immunotherapy.} 
	\label{Fig8}
\end{figure}

Simulation results show that on day $1$ without treatment, the tumor cell resistance metric follows a uniform distribution in $(0,1)$. After initiating treatment on day $5$, the peak of the resistance metric distribution gradually shifts rightward (Fig.\ref{Fig8}). This indicates that under drug selection pressure, most tumor cells develop increasing resistance, with the population evolving toward drug-resistant phenotypes. A comparison between Fig.\ref{Fig8}A and Fig.\ref{Fig8}B reveals more pronounced resistance evolution under targeted therapy. After targeted therapy, nearly no low-resistance subpopulations remain. In contrast, immunotherapy maintains drug-sensitive subpopulations until the end of the simulation. These results suggest stronger tumor cell sensitivity to targeted therapy, likely due to its direct action on specific molecular targets, which exerts stronger selection pressure. Conversely, immunotherapy indirectly eliminates tumor cells via immune activation, applying weaker selection pressure that permits persistence of sensitive subpopulations.

The findings based on model simulations provide crucial insights into the mechanisms of tumor drug resistance evolution under different treatment strategies. Simultaneously, they establish a theoretical foundation for optimizing combination therapies. For example, combining targeted therapy with immunotherapy may effectively delay the development of drug resistance. Targeted therapy directly eliminates tumor cells, while immunotherapy activates the immune system to remove residual resistant cells. Furthermore, for subpopulations with lower drug resistance, more precise treatment strategies can be designed. For instance, single-cell sequencing technology could identify characteristics of drug-sensitive cells to guide the development of targeted drugs or immunotherapies. Such precision treatment strategies may significantly improve efficacy while reducing side effects. In summary, these findings offer novel approaches to overcome tumor drug resistance and provide a scientific basis for developing personalized treatment strategies.

\subsection{Evaluation of treatment efficacy and immune cell distribution}
\label{Sec3.4}

\begin{figure}[htbp]
	\centering
	\includegraphics[width=16cm]{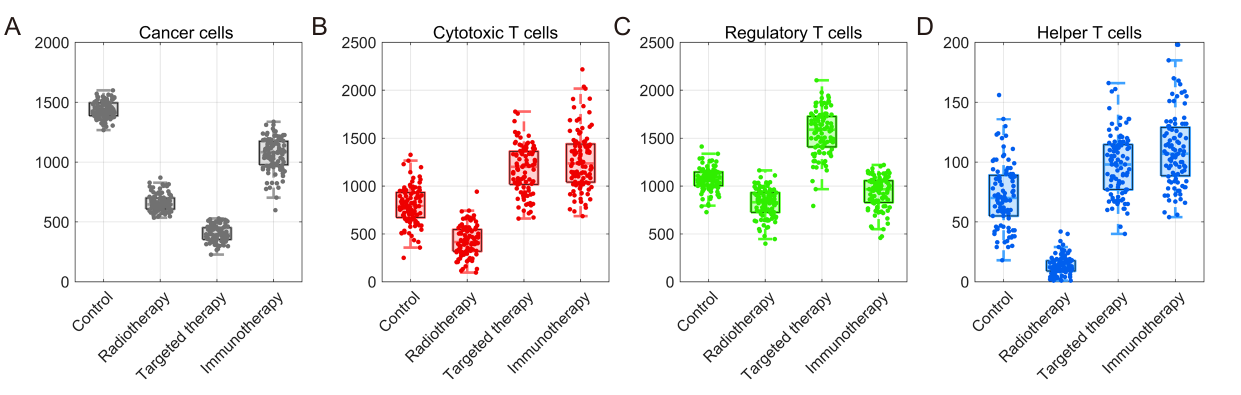}
	\caption{\textbf{Cell population comparisons under different treatment modalities at day $60$.} (A) Tumor cell count. (B) Cytotoxic T cell count. (C) Regulatory T cell count. (D) Helper T cell count.}
	\label{Fig9}
\end{figure}

To directly compare the therapeutic efficacy of radiotherapy, targeted therapy, and immunotherapy, we conduct $100$ stochastic simulations for the untreated group and three treatment groups. Figure \ref{Fig9} shows the boxplots of cell populations at day $60$ for each group. Results demonstrate that all three treatments significantly reduce tumor cell counts compared to untreated controls (Fig. \ref{Fig9}A). By day 60, tumor cell numbers decrease to $30\%$-$60\%$ of baseline values across treatments. Targeted therapy shows the strongest tumor growth suppression, followed by radiotherapy, then immunotherapy.

For immune cells, radiotherapy significantly reduces immune cell infiltration levels while inhibiting tumor cell proliferation (Fig.\ref{Fig9}B-D). In contrast, targeted therapy and immunotherapy promote immune cell proliferation. Notably, immunotherapy not only substantially increases CTL and Th cell proliferation but also effectively controls Treg expansion. This indicates that immunotherapy enhances the immune response in the tumor microenvironment by boosting effector T cell function while downregulating immunosuppressive cell activity. These results provide critical evidence for optimizing treatment strategies, suggesting that combination therapies may further improve therapeutic outcomes.

To comprehensively evaluate the efficacy of combination therapies, we perform $100$ stochastic simulations for each of three combinations: radiotherapy plus immunotherapy, radiotherapy plus targeted therapy, and targeted therapy plus immunotherapy. Each simulation runs for $60$ days with treatment initiation on day $5$. For quantitative assessment, we implement a tumor response classification system analogous to the Response Evaluation Criteria in Solid Tumors (RECIST 1.1) \cite{Eisenhauer.EJC.2009}, categorizing tumor response into four types based on tumor cell counts: progressive disease (PD), stable disease (SD), partial response (PR), and complete response (CR). The classification rules are detailed as:
\begin{equation}\label{assess_efficacy}
	\begin{cases}
		\mathrm{PD}, & C_{60} \geq (1 + 0.20) C_0, \\
		\mathrm{SD}, & (1 - 0.30) C_0 \leq C_{60} < (1 + 0.20) C_0, \\
		\mathrm{PR}, & (1 - 0.80) C_0 \leq C_{60} < (1 - 0.30) C_0, \\
		\mathrm{CR}, & C_{60} < (1 - 0.80) C_0.
	\end{cases}
\end{equation}
Here, $C_0$ denotes the tumor cell count at $t = 0$, and $C_{60}$ denotes the tumor cell count at $t = 60$.

\begin{figure}[htbp]
	\centering
	\includegraphics[width=14cm]{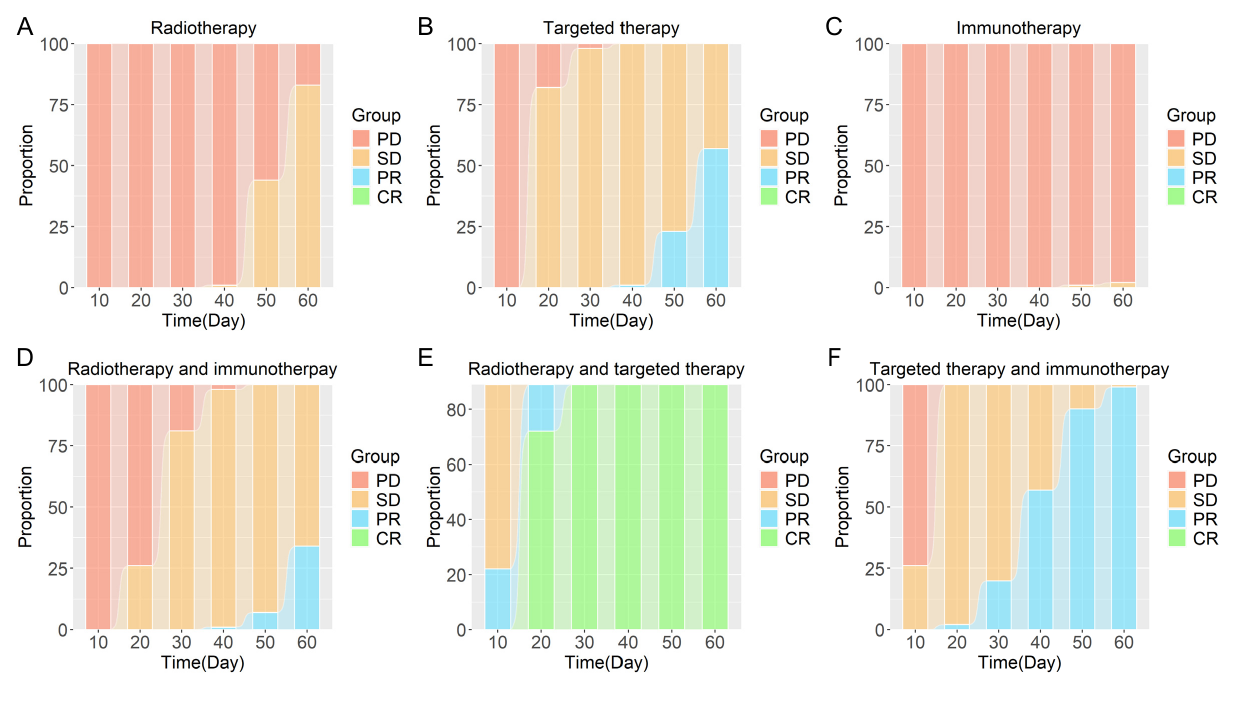}
	\caption{\textbf{Therapeutic efficacy evaluation under monotherapies and combination therapies.} (A) Radiotherapy. (B) Targeted therapy. (C) Immunotherapy. (D) Radiotherapy plus immunotherapy. (E) Radiotherapy plus targeted therapy. (F) Targeted therapy plus immunotherapy.}
	\label{Fig10}
\end{figure}

The simulation results demonstrate that all treatment modalities inhibit tumor progression to varying degrees (Fig.\ref{Fig10}). Among monotherapies, targeted therapy shows the fastest response, with approximately $80$\% of simulations achieving SD by day $20$, and over $50$\% reaching PR by the treatment endpoint (Fig.\ref{Fig10}B). Radiotherapy exhibits slower effects, with SD emerging at day $40$ and prevailing in most simulations (Fig.\ref{Fig10}A). Immunotherapy shows the most delayed response, with SD appearing only at day $50$ (Fig.\ref{Fig10}C). 

Among the combination therapies, radiotherapy plus targeted therapy shows the best efficacy (Fig.\ref{Fig10}E). Tumor growth is effectively controlled within $10$ days, with over $20$ simulations reaching PR directly (Fig.\ref{Fig10}E). By day $30$, all simulations achieve CR and maintain this status throughout the simulation. The other two combinations sequentially reached SD and ultimately PR (Fig.\ref{Fig10}D and F), with targeted therapy plus immunotherapy showing a faster onset.

Comparative analysis of all six strategies reveals significantly superior outcomes for combination therapies over monotherapies. This suggests that, when patient conditions permit, combination therapies can shorten treatment duration and substantially improve clinical outcomes. These findings provide critical evidence for optimizing clinical strategies and indicate broad application prospects for combination therapies in oncology.

\subsection{Tumor progression under different treatment strengths}
\label{Sec3.5}

\begin{figure}[htbp]
	\centering
	\includegraphics[width=16cm]{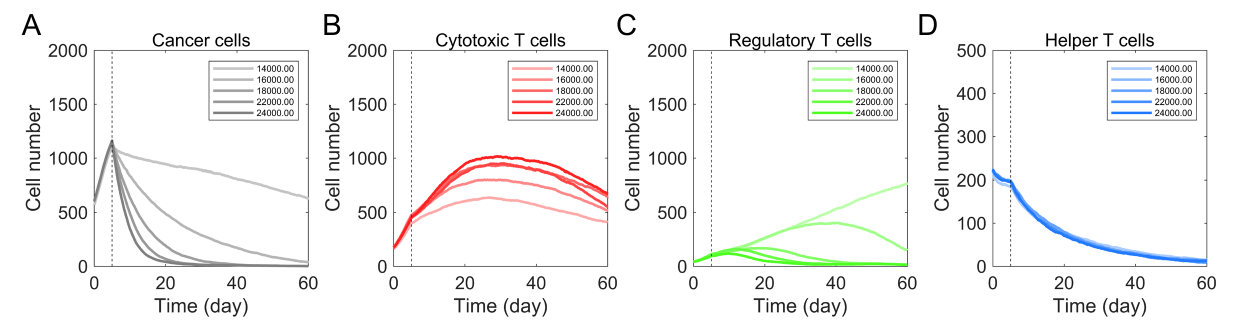}
	\caption{\textbf{Dynamic evolution of the tumor-immune system under varying radiotherapy intensities.} By adjusting the tumor cell radiosensitivity parameter $p_{dea, rad}^C$ to simulate different intensities---$p_{dea,rad}^C = 200.00$, $300.00$, $400.00$, $500.00$, $600.00$---five parameter sets are tested. (A) Tumor cells. (B) Cytotoxic T cells. (C) Regulatory T cells. (D) Helper T cells.}
	\label{Fig11}
\end{figure}

\begin{figure}[htbp]
	\centering
	\includegraphics[width=16cm]{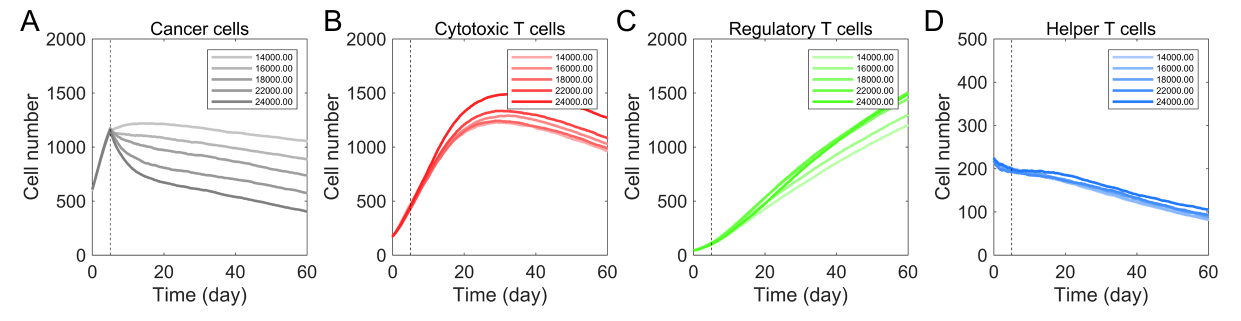}
	\caption{\textbf{Dynamic evolution of the tumor-immune system under varying targeted therapy intensities.} The regulatory coefficient $p_{dea, tar}^C$ for targeted therapy-induced tumor cell death is varied to simulate different treatment intensities, with five parameter values tested: 
$p_{dea,tar}^C = 200.00$, $300.00$, $400.00$, $500.00$, $600.00$. (A) Tumor cells. (B) Cytotoxic T cells. (C) Regulatory T cells. (D) Helper T cells.}
	\label{Fig12}
\end{figure}

\begin{figure}[htbp]
	\centering
	\includegraphics[width=16cm]{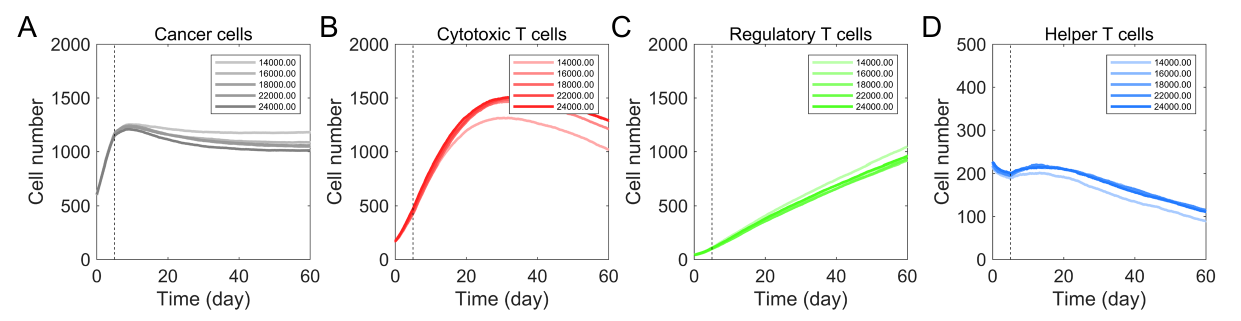}
	\caption{\textbf{Dynamic evolution of the tumor-immune system under varying immunotherapy intensities.} Different treatment intensities are simulated by adjusting the immunotherapy-induced tumor cell death rate coefficient $p_{dea, ICI}^C$, with five parameter sets: $p_{dea, ICI}^C = 14,000.00$, $16,000.000$, $18,000.00$, $22,000.00$, $24,000.00$. (A) Tumor cells. (B) Cytotoxic T cells. (C) Regulatory T cells. (D) Helper T cells.}
	\label{Fig13}
\end{figure}

To further investigate the dynamic evolution of the tumor-immune system under different treatment intensities, we adjust three key therapeutic parameters: the radiotherapy-induced death rate coefficient $p_{dea, rad}^C$, the targeted therapy-induced death rate coefficient $p_{dea, tar}^C$, and the immunotherapy-induced death rate coefficient $p_{dea, ICI}^C$. Each treatment strategy is tested at five intensity levels, with $20$ independent replicates per level. Treatment efficacy is evaluated by analyzing averaged cell population dynamics (Figs. \ref{Fig11}--\ref{Fig13}). This approach enables quantitative comparison of therapeutic efficacy across strategies and their dynamic impacts on the TME.

The simulation results show that after radiotherapy initiation, tumor cell depletion accelerates significantly with increasing treatment intensity (Fig.\ref{Fig11}A). When the tumoricidal efficacy parameter $p_{dea, rad}^C$ increases from $200.00$ to $300.00$, tumor cell numbers decrease substantially, but further increases in $p_{dea, rad}^C$ yield diminishing returns. For immune cells, higher radiotherapy intensity markedly elevates peak CTL counts (Fig.\ref{Fig11}B), while Th cell numbers remain largely unaffected (Fig.\ref{Fig11}D). Thus, moderate intensity enhancement improves both tumor suppression and immune response in the TME. However, when $p_{dea, rad}^C$ is increased from $400.00$ to $600.00$, complete tumor eradication occurs by day $35$ with parallel elimination of Tregs (Fig.\ref{Fig11}C), showing nearly identical depletion curves. These results suggest that maintaining $p_{dea, rad}^C$ below $400.00$ preserves normal immune function while avoiding therapeutic saturation and excessive patient burden.

Following targeted therapy initiation, tumor cell counts decrease consistently with increasing treatment intensity. Specifically, each $100.00$ increment in $p_{dea, tar}^C$ reduces final tumor cell counts by approximately $170$, demonstrating effective tumor suppression (Fig.\ref{Fig12}A). Concurrently, immune cell populations increase variably with treatment intensity (Fig.\ref{Fig12}B-D). Notably, when $p_{dea, tar}^C$ reaches $600.00$, CTL proliferation significantly increases. This indicates that while targeted drugs directly kill tumor cells without explicitly regulating immune cell proliferation or apoptosis, they indirectly remodel the tumor microenvironment to enhance antitumor immunity.

Increasing immunotherapy intensity enhances tumor growth inhibition (Fig.\ref{Fig13}A). Comparative analysis of tumor and immune cell populations reveals more pronounced effects on immune cells (Fig.\ref{Fig13}B-D), particularly the significant increase in CTL numbers (Fig.\ref{Fig13}B), demonstrating immunotherapy primarily suppresses tumors by improving immune activity in the TME. Compared with Fig. \ref{Fig7}B, Fig. \ref{Fig13}D more clearly displays the enhancing effect of immunotherapy on Th cell numbers. Notably, when $p_{dea, ICI}^C$ ranges from $18,000.00$ to $24,000.00$, Treg and Th cell counts stabilize over time (Fig.\ref{Fig13}C-D), suggesting an intensity threshold beyond which dose escalation may not further improve efficacy. These results indicate that optimal dosing should balance therapeutic effects and potential adverse events to avoid overtreatment.

\section{Conclusions}
\label{Sec4}

In this study, we developed an agent-based, two-dimensional tumor-immune model that incorporates cellular heterogeneity as a key determinant of cell behavior and intercellular interactions. This model computationally simulates the spatiotemporal evolution of tumor-immune system interactions under various treatments. We first explored the natural evolution patterns of untreated tumor microenvironments. We then systematically evaluated monotherapies and combination therapies. Further investigations revealed the dynamic evolution of drug resistance during treatment. These results not only elucidate resistance mechanisms but also quantify the relationship between treatment intensity and therapeutic efficacy, providing critical theoretical foundations for optimizing clinical regimens and overcoming resistance.

Extensive studies have demonstrated that the immune system plays a pivotal role in tumor development and treatment \cite{deVisser.CancerCell.2023}. Previous research has employed multiscale modeling to investigate tumor-immune interactions under different therapeutic strategies \cite{Anderson.Cell.2006,Zhang.MathComputSimul.2009,Gong.JRSocInterface.2017,Jalalimanesh.MathComputSimul.2017,Hickey.CellSystems.2024}, highlighting the crucial role of agent-based models in capturing tumor evolution dynamics. However, these studies have three main limitations: (1) Oversimplified immune cell configurations that lack immunosuppressive components such as Tregs; (2) Insufficient consideration of how cellular heterogeneity affects cell-cell interactions; (3) A predominant focus on single-drug resistance mechanisms with limited systematic comparison of treatment strategies.

To address these gaps, our model integrates cellular heterogeneity and diverse immune cell types to more realistically simulate the tumor microenvironment, providing a comprehensive evaluation platform for studying resistance evolution across different therapeutic regimens.

Nonetheless, the current model has several limitations. First, it simplifies tumor-immune interactions by focusing solely on cellular-scale dynamics, omitting molecular-scale components and the complex regulatory roles of key cytokines in the tumor microenvironment. Second, the model lacks integration with clinical data and does not include parameter calibration for specific cancer types, which limits its direct clinical applicability. Effectively integrating experimental data into agent-based models remains a fundamental challenge in tumor modeling. Recent research has explored strategies to improve data-model integration to enhance clinical relevance. For example, multiplex imaging has been used to incorporate spatial omics data into agent-based models to study how tumor phenotypic transitions affect T cell efficacy \cite{Hickey.CellSystems.2024}. Additionally, due to computational resource constraints and model complexity, our simulations are currently limited to two-dimensional space, whereas three-dimensional modeling would more accurately reflect in vivo tumor biology.
 
To address these limitations, we propose four directions for future model optimization: 
\begin{enumerate}
\item[(1)] Multiscale integration---incorporate differential equation models of key cytokines and signaling pathways at the molecular level to refine microenvironment regulation;
\item[(2)] Drug mechanism incorporation---integrate pharmacokinetic/pharmacodynamic (PK/PD) models to precisely simulate drug absorption, distribution, and metabolism;
\item[(3)] Data-driven refinement---utilize single-cell RNA sequencing, spatial transcriptomics, and clinicopathological data to enhance parameter realism and improve predictive accuracy;
\item[(4)] Computational upgrade---develop GPU-accelerated, three-dimensional agent-based modeling platform with parallel computing capabilities and interaction optimization for large-scale spatial simulations. 
\end{enumerate}
These improvements will significantly enhance the model's biological fidelity and computational performance, providing a more powerful tool for mechanistic investigation and clinical translation of cancer immunotherapy.

\section*{Author contributions}

\textbf{Yuhong Zhang:} Conceptualization; Data curation; Investigation; Methodology; Software; Validation; Visualization; Writing – original draft, Writing review \& editing. \textbf{Chenghang Li:} Conceptualization; Methodology; Visualization; Writing – original draft, Writing review \& editing. \textbf{Boya Wang:} Visualization; Data curation; Software. \textbf{Jinzhi Lei:} Conceptualization; Formal analysis; Funding acquisition; Investigation; Methodology; Project administration; Resources; Supervision; Writing – original draft, Writing review \& editing.

\section*{Acknowledgments}
This work is supported by the National Natural Science Foundation of China (NSFC 12331018).

\section*{Declaration of interest}
The authors declare no competing interests.

\section*{Code availability}
The codes used in this study are available at https://github.com/jinzhilei/ABM-Tumor-Immune-Dynamics.

\bibliographystyle{elsarticle-num}
\bibliography{references.bib}

\end{CJK}
\end{document}